\documentclass[fleqn,usenatbib]{mnras}

\usepackage{newtxtext,newtxmath}

\usepackage[T1]{fontenc}
\usepackage{ae,aecompl}

\usepackage{graphicx}	
\usepackage{amsmath}	
\usepackage{amssymb}	
\usepackage{scalerel}   
\usepackage{booktabs}
\usepackage{multirow}
\usepackage{enumitem}
\setlist[enumerate]{leftmargin=*}
\usepackage[dvipsnames]{xcolor}
\usepackage{siunitx}

\hypersetup{final}

\title[Observing early phase SNe with strong lenses]{Observing the earliest moments of supernovae using strong gravitational lenses}

\author[M. Foxley-Marrable et al.]{Max Foxley-Marrable,$^{1}$\thanks{E-mail: max.foxley-marrable@port.ac.uk}
Thomas E. Collett,$^{1}$
Chris Frohmaier,$^{1}$
\newauthor Daniel A. Goldstein,$^{2,4}$
Daniel Kasen,$^{3}$
Elizabeth Swann$^{1}$
and
David Bacon$^{1}$\\
$^{1}$Institute of Cosmology and Gravitation, University of Portsmouth, Dennis Sciama Building, Burnaby Road, Portsmouth, PO1 3FX, UK\\
$^{2}$California Institute of Technology, 1200 East California Blvd, MC 249-17, Pasadena, CA 91125, USA \\
$^{3}$Department of Astronomy, University of California, Berkeley, 501 Campbell Hall, Berkeley, CA 94720, USA \\
$^{4}$Hubble Fellow
}

\date{Accepted 2020 May 3. Received 2020 April 25; in original form 2020 March 13. }

\pubyear{2019}

\begin{document}
\label{firstpage}
\pagerange{\pageref{firstpage}--\pageref{lastpage}}
\maketitle

\begin{abstract}
{We determine the viability of exploiting lensing time delays to observe strongly gravitationally lensed supernovae (gLSNe) from first light}. Assuming a plausible discovery strategy, the Legacy Survey of Space and Time (LSST) and the Zwicky Transient Facility (ZTF) will discover $\sim$ 110 and $\sim$ 1 systems per year before the supernova (SN) explosion in the final image respectively. Systems will be identified $11.7^{+29.8}_{-9.3}$ days before the final explosion. {We then explore the possibility of performing early-time observations for Type IIP and Type Ia SNe in LSST-discovered systems.} Using a simulated Type IIP explosion, we predict that the shock breakout in one trailing image per year will peak at $\lesssim$ 24.1 mag ($\lesssim$ 23.3) in the $B$-band ($F218W$), however evolving over a timescale of $\sim$ 30 minutes. Using an analytic model of Type Ia companion interaction, we find that in the $B$-band we should observe at least one shock cooling emission event per year that peaks at $\lesssim$ 26.3 mag ($\lesssim$ 29.6) assuming all Type Ia gLSNe have a 1 M$_\odot$ red giant (main sequence) companion. {We perform Bayesian analysis to investigate how well deep observations with 1 hour exposures on the European Extremely Large Telescope would discriminate between Type Ia progenitor populations.} We find that if all Type Ia SNe evolved from the {double-degenerate} channel, then observations of the lack of early blue flux in 10 (50) trailing images would rule out more than 27\% (19\%) of the population having 1 M$_\odot$ main sequence companions at 95\% confidence.
\end{abstract}



\section{Introduction} \label{sec: Intro}
Early observations of supernovae (SNe) light curves are critical in constraining the properties of SN progenitor systems and their pre-explosion evolution in a way that cannot be inferred from late-time observations (e.g. \citealt{Kasen2010ApJ, Piro2010, Rabinak2011, Piro2016, Kochanek2019, Fausnaugh2019, Yao2019, Miller2020, Bulla2020}). In addition, the physics of SN explosion mechanisms are still yet to be well understood (see \citealt{Smartt2009, Janka2012, Hillebrandt2013, Burrows2013, Maoz2014, Livio2018} for recent reviews). 

The earliest expected SN emission should comprise of a bright X-ray/UV flash as the initial radiation-mediated shock propagates to the outer edges of the star, ejecting the envelope in a process known as the `shock breakout' (see \citealt{Colgate1968, Colgate1975, Grassberg1971, Lasher1975, Lasher1979, Imshennik1977, Falk1978, Klein1978, Epstein1981, Ensman1992, Piro2010}). This process occurs over a timescale of order seconds to a fraction of an hour, dependent on the progenitor size. If there is significant circumstellar material surrounding the progenitor prior to the moment of explosion, the breakout timescale could be extended to a number of days. {After the initial shock breakout, as the ejected envelope expands, we expect to observe UV/optical cooling emission evolving over a timescale of order days} (see \citealt{Waxman2017} and references therein for a comprehensive theoretical overview on the topic of shock breakout and cooling emission). \newpage

The progenitors of Type Ia SNe remain an unsolved problem in astrophysics \citep{Maoz2014, Livio2018}, with the single-degenerate (SD) and double-degenerate (DD) channels being plausible explanations for the post-explosion light curves. The SD scenario occurs when a carbon/oxygen (C/O) white dwarf (WD) accretes mass from a non-degenerate companion star, triggering an explosion via thermonuclear detonation on the approach to the Chandrasekhar Mass, $M_\mathrm{ch}$ \citep{WhelananfIben1973ApJ, nomoto1982accreting, Maguire2017}. In the DD scenario, a WD approaches $M_\mathrm{ch}$ due to accretion of mass or directly merging with a secondary WD \citep{IbenandTutukov1984ApJ, Webbink1984ApJ, Maguire2017}. {Another plausible model is the sub-$M_\mathrm{ch}$ `double-detonation' scenario, where an initial detonation in the outer helium layers accreted onto the surface of the WD triggers a secondary detonation in the C/O core \citep{Nomoto1980, Taam1980, Woosley1986, Livne1990, Woosley1994, Fink2010, Moll2013}}. This mechanism has been used to explain the unusual colour evolution and spectra of three recent Type Ia SN \citep{Jiang2017, De2019, Jacobson-Galan2019}. 

Type Ia SNe are used to measure cosmological distances \citep[e.g.][]{Riess1998, Perlmutter1999} on the assumption their peak magnitudes are all effectively homogeneous after standardisation ($\sigma_M \sim 0.1$ mag; e.g. \citealt{Betoule2014, Macaulay2017, Jones2018}). Therefore, if the mean intrinsic brightness of the Type Ia SN significantly varies with progenitor model, and the progenitor population varies with redshift \citep{Childress2014}, cosmological analyses dependent on SNe Ia will be inherently biased. Since neither the SD or DD channels have been ruled out conclusively, it is entirely plausible that both scenarios are valid, and that traces of the population could even come from other channels {\citep[e.g. the core-degenerate channel, see][and references therein]{Livio2018}}. {Early photometry obtained within hours or days of the SN Ia explosion could provide insight into the presence of a potential companion star and constrain properties such as the companion radius \citep[e.g.][]{Nugent2011, Bloom2012, Goobar2014, Goobar2015, Marion2016, Hosseinzadeh2017, Dimitriadis2019, Shappee2019ApJ}}.

Even with the development of wide-field optical surveys, observing the earliest moments of SNe is still non-trivial and heavily reliant on chance. Ideally, we would like to systematically predict the precise moment a SN will appear on a particular patch of sky and start performing high-cadence observations in the moments prior to and at first light. Such a prediction could be possible if the SN was subject to strong gravitational lensing \citep{Suwa2018}.

Gravitational lensing occurs because massive objects e.g. elliptical galaxies, deform the local curvature of spacetime such that nearby rays of light become deflected from their original path. When a sufficiently dense object is precisely aligned between the observer and a background source, multiple images of the background object form. This effect is known as strong gravitational lensing \citep{Einstein1936, Zwicky1937}. The light travel time from the source to the observer varies between lensed images due to geometrical differences in the path length and differences in gravitational time dilation. Both effects are a function of the path of the light through the gravitational potential of the lens \citep{Shapiro1964PhRvL..13..789S, BlandfordandNarayan1986ApJ, Treu2016}.

When a strongly lensed supernova (gLSN) explodes, an observer will witness the SN from first light once in each lensed image, but with a time delay between the images. Hence, if a gLSN is identified before the appearance of the SN in any of the multiple images, and the mass distribution of the lens is well understood, it should be possible to predict the explosion time of the SN in the remaining images. 

\begin{table}
    \centering
    \begin{tabular}{lllll}
    \toprule
    \multirow{2}{*}{SN Type} &
    \multicolumn{2}{c}{LSST} & 
    \multicolumn{2}{c}{ZTF}  \\
    & Doubles & Quads & Doubles & Quads \\
    \midrule
    
    IIn  &  52.0  &  9.7  &  0.1  &  0.5  \\
    IIP  &  18.9  &  3.3  &  0.2  &  0.1  \\
    Ia   &  12.8  &  1.5  &  ---  &  0.1  \\
    Ibc  &   3.4  &  0.9  &  ---  &  ---  \\
    IIL  &   2.2  &  0.8  &  ---  &  ---  \\
    91T  &   1.6  &  0.2  &  ---  &  ---  \\
    91bg &   0.2  &  0.1  &  ---  &  ---  \\
    
    \bottomrule
    
    Total & 91.1  & 16.5  &  0.3  &  0.7
    \end{tabular}
    \caption{Number of gLSNe discovered with one or more unexploded trailing images each year. Rates below 0.05 per year are not shown.}
    \label{tab:noPerYear}
\end{table}

\begin{table*}
    \centering
    \begin{tabular}{l|l}
        \toprule
        Subplot & Description \\
        \midrule
        (a) - (c) & Observer frame apparent magnitudes for trailing lensed images in the $g$, $r$ and $i$ bands respectively.  \\
        (d) & Redshift of the background source. \\
        (e) & `Reaction' time between the discovery and confirmation of the gLSNe and the appearance of the final image. \\
        (f) & Error in the time delay relative to the first image for all trailing images.\\
        \bottomrule
    \end{tabular}
    \caption{Description of subplots for {Figures} \ref{fig:superMegaLSST}-\ref{fig:superMegaZTF} and {Figures} \ref{fig:superMegaLSSTIIn}-\ref{fig:superMegaLSST91bg} in the appendix.}
    \label{tab:superMegaFigDescriptions}
\end{table*}

SN Refsdal, a core-collapse SN multiply imaged by a foreground galaxy cluster \citep{Kelly2015}, was predicted to have a fifth image appear $\sim$ 1 year from the appearance of the first image \citep{Treu2016Refsdal}. This prediction was later confirmed by the reappearance of the SN in the fifth lensed image \citep{Kelly2016}. However, the errors on the predictions ranged from ~5-20\% of the year-long time delay between the first and fifth image, dependent on the choice of lens model \citep{Treu2016Refsdal}. This can be attributed to the dense and complicated mass profile of the foreground galaxy cluster lens. Therefore lens systems with significantly simpler mass profiles (e.g. elliptical galaxies) and shorter time delays are more suited for very early observations of lensed SN light curves.

To date, only one other gLSN with resolved images has been discovered \citep[iPTF16geu,][]{Goobar2017}, and this was identified after the appearance of the last image. A sample of gLSN with followup triggered before the reappearance of the SN in the remaining images is required to constrain progenitor populations. The Legacy Survey of Space and Time (LSST) and Zwicky Transient Facility (ZTF) are the next generation of wide-field, high-cadence imaging surveys which together are expected to yield thousands of gLSNe over the next decade \citep{Goldstein2018ApJ, GoldsteinNugentGoobar2019}.

In this paper, we endeavour to answer the following questions:
\begin{enumerate}
    \item Will LSST and ZTF enable the discovery of gLSNe before the appearance of all multiple images?
    \item How long is the time frame between the discovery of the system and explosion of the last image? How precisely can the last explosion time be predicted?
    \item How bright will the early phase light curves of Type IIP and Type Ia SNe found in the trailing images of LSST-discovered gLSNe get?
    \item Can we use LSST-discovered gLSNe to make inferences on the progenitor population of Type Ia SNe with redshift? How will this compare with constraints from unlensed SNe Ia?
    \item Can we measure precise time delays between the rapid early-phase light curves of gLSNe?
\end{enumerate}

In Section \ref{sec: Populations}, we use the gLSNe catalogues from \cite{GoldsteinNugentGoobar2019} to provide predictions into the populations of gLSNe in LSST and ZTF that will be discovered before reappearance of the SN explosion in any of the remaining lensed images. In Section \ref{sec: earlyPhaseSNe}, we make predictions on the magnitude distributions of early phase, LSST-discovered Type IIP and Type Ia SNe, whose light curves were generated using the SuperNova Explosion Code (SNEC) code and the companion emission models of \cite{Kasen2010ApJ}, respectively. In Section \ref{sec: progenitorpop} we explore how gLSNe can be used to constrain SNe Ia populations. In Section \ref{sec: cosmography} we determine whether early phase SNe observations are useful for the field of time delay cosmography. We then conclude in Section \ref{sec: Conc}.

\begin{figure*}
    \centering
    \includegraphics[width=0.9\textwidth]{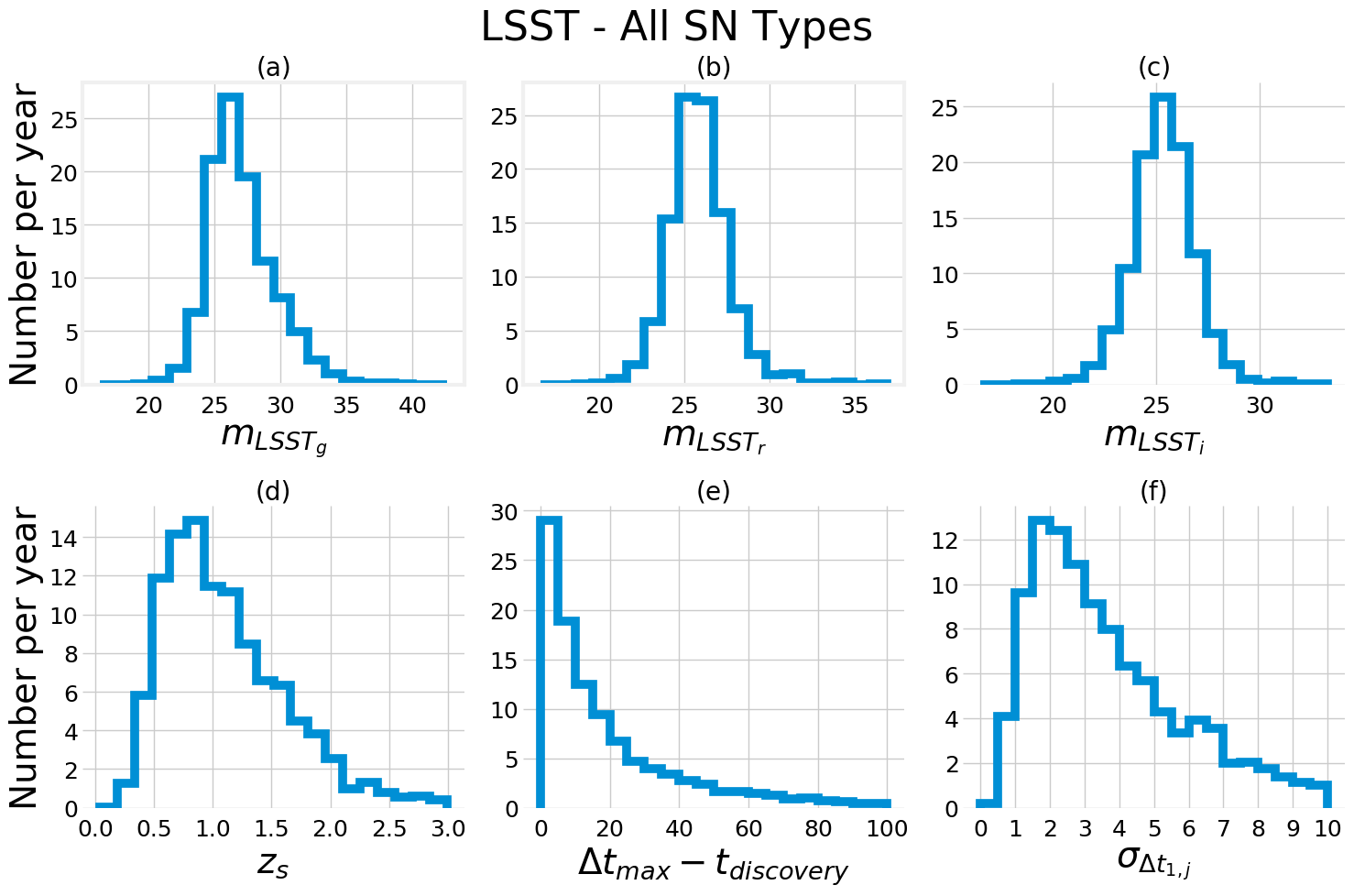}
    \caption{{Distributions and annual rates of LSST-discovered gLSNe (of all SN Types) containing trailing images with unexploded SNe. See Table \ref{tab:superMegaFigDescriptions} for descriptions of the subplots.}}
    \label{fig:superMegaLSST}
\end{figure*}

\begin{figure*}
    \centering
    \includegraphics[width=0.9\textwidth]{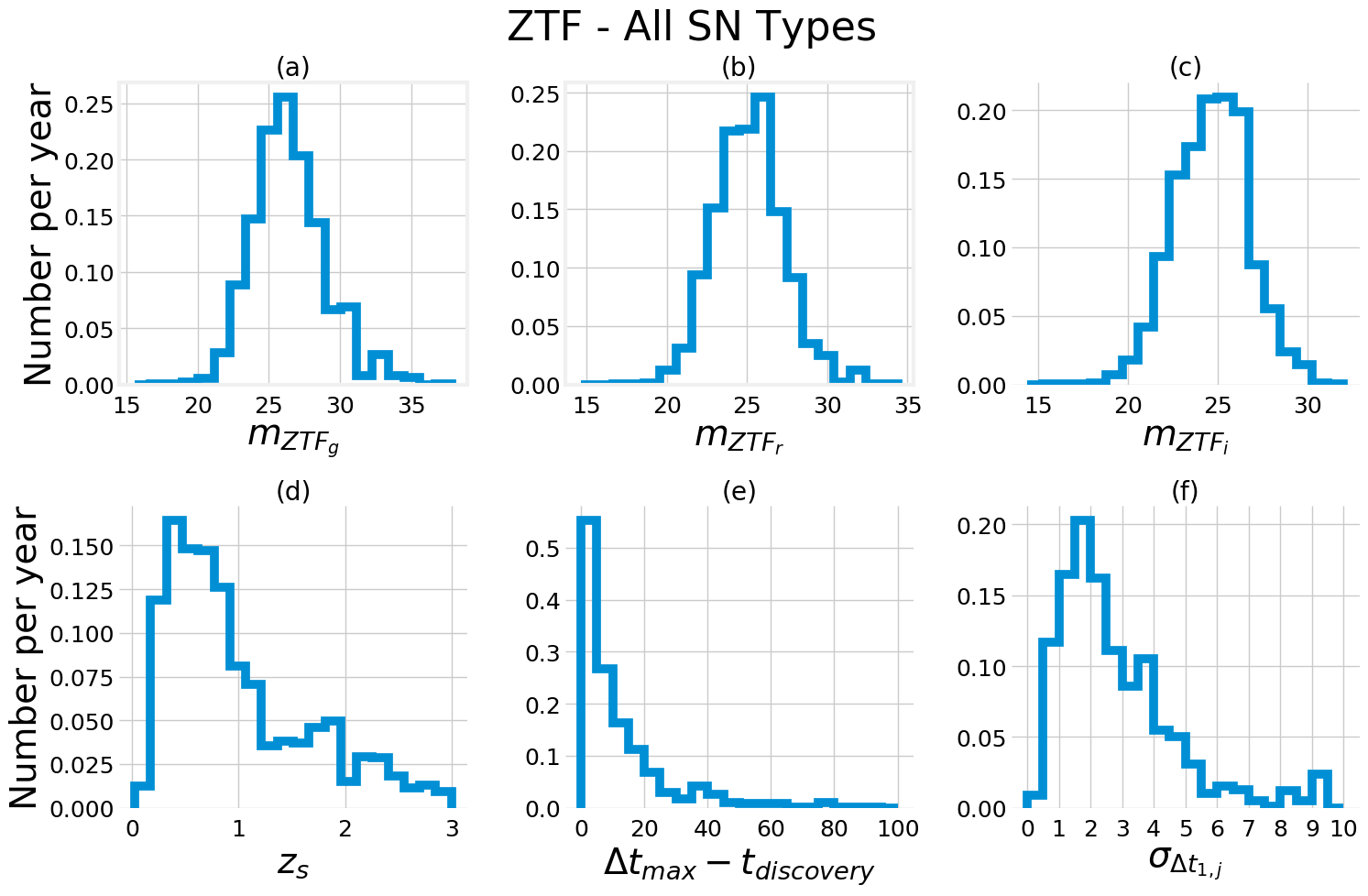}
    \caption{{Distributions and annual rates of ZTF-discovered gLSNe (of all SN Types) containing trailing images with unexploded SNe. See Table \ref{tab:superMegaFigDescriptions} for descriptions of the subplots.}}
    \label{fig:superMegaZTF}
\end{figure*}

\begin{figure*}
    \centering
    \includegraphics[width=\textwidth]{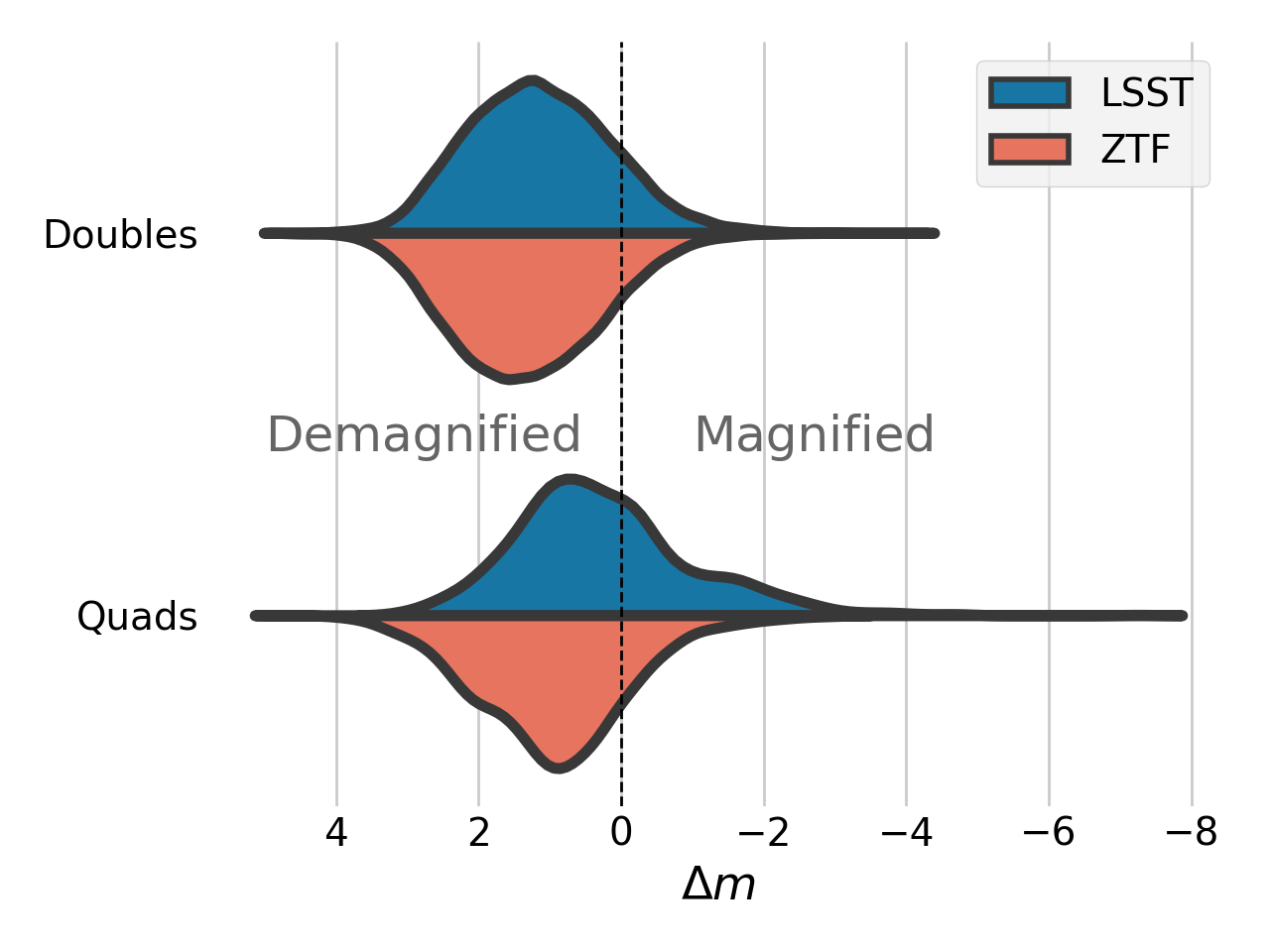}
    \caption{{Distribution of trailing image magnifications (shown as deviation in magnitudes, $\Delta m$) for double and quadruple image gLSNe after discovery by LSST and ZTF.}}
    \label{fig:deltamagImgno}
\end{figure*}

\section{LSST/ZTF Populations} \label{sec: Populations}
\subsection{gLSN Catalogues}
To make predictions on the populations of LSST/ZTF discovered gLSNe with `trailing' SN images, i.e. gLSNe discovered before the reappearance of the SN in the remaining lensed images, we use the publicly available simulated gLSN catalogues from \cite{GoldsteinNugentGoobar2019}\footnote{\texttt{https://portal.nersc.gov/project/astro250/glsne/}}. These catalogues were created by simulating a population of randomly realised gLSNe systems into mock LSST/ZTF survey data and applying the resolution-insensitive discovery strategy detailed in Section 4.1 of \cite{Goldstein2018ApJ} to forecast the properties and rates of gLSNe to be discovered by LSST and ZTF. 

Only elliptical galaxies were considered as potential lenses in the catalogues. Ellipticals are the most common type of gravitational lens; the sharp 4000 \r{A} break in their uniform spectra allows their photometric redshifts to be accurately measured; and they are the only lens compatible with the \citet{Goldstein2018ApJ} discovery strategy. The projected mass distributions of the ellipticals were modelled as a singular isothermal ellipsoids \citep{Kormann1994}, shown to be in good agreement with observations \citep[e.g.][]{Koopmans2009}.

The catalogues contain 7 different subtypes of gLSNe: including 3 subtypes of thermonuclear gLSNe (Type Ia, SN 1991bg-like and SN 1991T-like), with rates and luminosity functions based on \citet{Sullivan2006}; and 4 subtypes of core-collapse gLSNe (Type IIP, Type IIL, Type IIn, Type Ib/c) with rates and luminosity functions based on \citet{Li2011MNRAS}. The rates in the gLSN catalogue carry uncertainties of order 10\% which carries over to the rates presented in our analysis. Three different types of host galaxies were considered in the catalogues: elliptical galaxies (very little to no star formation), S0/a-Sb galaxies (some star formation) and late-type/spiral galaxies (ongoing star formation). The simulations assume elliptical and S0/a-SB galaxies only host normal SNe Ia and SN 1991bg-like events, whereas late-type/spiral galaxies host both core-collapse and thermonuclear SNe types.

With the assumptions listed above, for each gLSN system the properties of the lens galaxy, the SN and the host galaxies were realised at random, uniformly distributed on the sky and assigned a reddening value $E(B-V)$ for the host galaxy and {Milky Way} dust\footnote{Lens galaxy dust was neglected.}.

For ZTF, \citet{GoldsteinNugentGoobar2019} used the simulated survey data and scheduler from \citet{Bellm2019} for the public, partnership and Caltech programs. For LSST, both the \texttt{minion1016} \citep{Delgado2014} and \texttt{altsched} \citep{Rothchild2019} observing strategies were considered. For our analysis we only consider the \texttt{altsched} observing strategy\footnote{Yields are comparable to \texttt{minion1016}, but with better sampled light curves that are discovered earlier.}.

\subsubsection{Discovery Strategy}
The discovery strategy proposed in \citet{Goldstein2018ApJ, GoldsteinNugentGoobar2019} is designed to photometrically identify gLSNe in transient survey data without the need to resolve the multiple images through follow-up observations. 

The discovery strategy can be summarised as follows: first, identify SNe candidates spatially aligned with elliptical galaxies. Since there is very little to no ongoing star formation in elliptical galaxies, they primarily host only Type Ia SNe \citep{Li2011MNRAS}. The next step is to test whether the SNe candidate is a Type Ia SN hosted by the elliptical galaxy. This can be achieved by comparing the properties of the SN light curve (e.g. peak brightness, light curve shape and colour evolution) to a SN Ia template \citep[e.g. SALT2;][]{Guy2007} at the photometric redshift of the elliptical galaxy. If observations are inconsistent with a SN Ia at the photometric redshift of the apparent host, then it is a candidate for strong lensing. A transient is identified as a gLSN when at least one data point is observed with a 5$\sigma$ discrepancy from the best fit Ia light curve (consistent with the elliptical's photometric redshift) and at least four data points have signal-to-noise $\geq$ 5 (see Section 4.2 of \citealt{Goldstein2018ApJ}).

\subsection{Trailing gLSNe Populations}
A system in the gLSNe catalogue is determined to contain unexploded trailing images if the arrival time of any lensed image is after the discovery time of the gLSN. The moment of explosion for each image is calculated by adding the time delay to the arrival time of the first image at zero-phase, and subtracting the difference between explosion time and zero-phase time for each model. For Type Ia and Type IIP SNe, the zero-phase time, $t_0$ is at the peak of the SN light curve, for the other models, $t_0$ is the explosion time. To determine the time of explosion, we assume the explosion time to be 20 rest frame days before peak for Type Ia SNe and 19 rest frame days for Type IIP SNe. This is derived from the difference between peak and the earliest non-zero data point of the \cite{Hsiao2007ApJ_Ia} and \cite{Sako2011ApJ_IIP} models respectively.

The populations of gLSNe with unexploded trailing images for all SN types are illustrated in Figures \ref{fig:superMegaLSST} and \ref{fig:superMegaZTF} (see Figures \ref{fig:superMegaLSSTIIn} - \ref{fig:superMegaLSST91bg} for a breakdown of the LSST distributions by SN Type\footnote{ZTF distributions were purposefully left out due to low sample size, resulting in distributions being dominated by shot noise.}). The number of discoveries per year for each instrument and SNe type are shown in Table \ref{tab:noPerYear}. Across all SN types LSST is expected to find $\sim$ 110 trailing gLSNe per year, whilst ZTF will yield a significantly lower rate of systems at $\sim$ 1 trailing gLSNe per year.

The ZTF sample is dominated by quadruple imaged systems (hereby referred to as `quads') whereas the LSST sample is dominated by double imaged systems (hereby referred to as `doubles'). Quads dominate the ZTF sample because ZTF is shallow and quads typically have higher magnification than doubles. The deeper, lower cadence of LSST allows it to find fainter systems but at later times: since doubles typically have longer time delays than quads, they are more likely in LSST. Quads make up $\sim$ 16 \% of the total sample with a discovery rate of approximately once every 1.4 years with ZTF and once every 22 days with LSST. Across all quadruple gLSNe types, we expect to find $\sim$ 15 per year in LSST and $\sim$ 1 per year in ZTF with a single trailing image remaining. This falls to $\sim$ 1 quad per year in LSST ($<$ 0.01 in ZTF) with 2 or more images remaining. In many ways, quads are more suited for early phase SN observations, since they are typically more highly magnified, and they are easier to accurately model enabling more precise predictions of the time delay. However, the shorter time delays make the rate of quads discovered before the final explosion far lower than the double systems.

Lens modelling of galaxy scale lenses typically yields model time delay estimates at around 5\% precision \citep[e.g.][]{Wong2017_H0LiCOWIV, Birrer2019_H0LiCOWIX, Chen2019_H0LICOW}. We assume this fractional precision for the predicted reappearance of trailing images. Comparable fractional precision was achieved for predicting the reappearance of SN Refsdal in a much more complicated cluster lensing environment \citep{Treu2016Refsdal}. Galaxy scale lenses should be easier to precisely model, though the shorter time delay will require a fast turn around between discovery and time delay estimate. Assuming this 5\% error is achieved for incomplete systems we find that typically we will be able to predict the time delays to $3.2_{-1.6}^{+3.4}$ days around the appearance of the final image. Very few trailing images are predictable to less than a day (Figures \ref{fig:superMegaLSST}f and \ref{fig:superMegaZTF}f). The `reaction' time between discovery of the gLSN and the appearance of the SN in the final lensed image (Figures \ref{fig:superMegaLSST}e and \ref{fig:superMegaZTF}e) is typically within $11.7^{+29.8}_{-9.3}$ days from discovery. Performing follow-up observations and modelling lenses within this time scale will pose a challenge, however the promise of automated lens modelling software (e.g. \texttt{AutoLens}; \citealt{Nightingale2018}) could alleviate this time pressure.

The trailing images in the gLSNe sample have a peak median magnitude of $25.4_{-1.3}^{+1.4}$ in the $i$-band, which is typically dimmer than the unlensed SN explosion (see Figures 11 and 25 of \citealt{GoldsteinNugentGoobar2019} for comparison). This is due to the vast majority of trailing gLSNe only having one image remaining after discovery, which are commonly demagnified by $\sim$ 1 or 2 magnitudes (see Figure \ref{fig:deltamagImgno}). {This is because the final image is typically closest to the centre of the lensing galaxy. These images have significant mass density at their location, such that the light rays are over-focused. Small changes in the image plane position result in large changes in the source plane position, so these images are demagnified.} Coupled with extinction by dust, it is clear that obtaining early phase SN data from the trailing images of gLSNe will be an observationally expensive effort. 

\subsection{Unknown vs Known Lenses}

\begin{table}
    \centering
    \begin{tabular}{llll}
    \toprule
    \multirow{2}{*}{SN Type} &
    \multicolumn{2}{c}{$\Delta$N} &
    \multirow{2}{*}{$\Delta$m} \\
     &  Doubles & Quads & \\
    \midrule
    IIn   &  24.4 & 6.1  &  -0.4  \\
    IIP   &  8.7  & 2.7  &  -0.4  \\
    Ia    &  4.5  & 1.1  &  -0.3  \\
    Ibc   &  1.2  & 0.4  &  -0.2  \\
    IIL   &  1.0  & 0.5  &  -0.2  \\
    91T   &  0.6  & 0.1  &  -0.2  \\
    91bg  &  ---  & --- &   -0.5  \\
    \bottomrule
    Total &  40.4 & 10.9 &
    \end{tabular}
    \caption{Change in the number and average brightness of gLSNe with trailing images if all lensed SNe in the LSST catalogue were already known lenses. In this scenario, discovery is assumed from the first observation of the SN with a signal-to-noise $\geqslant$ 5 in any filter. Rates below 0.05 per year are not shown.}
    \label{tab:unknown_to_known}
\end{table}

Our estimated yields are potentially pessimistic, since the assumed discovery method does not include the possibility that the SN host is already known to be strongly lensed. LSST is expected to discover $\sim$ 100,000 lenses \citep{Collett2015} and immediate followup of any transient detected in a known lens system should enable the identification of gLSN at an earlier phase than we have assumed. {For LSST-discovered gLSNe, by assuming that all lenses in the \citet{GoldsteinNugentGoobar2019} catalogue are already known} and assuming gLSN discovery from the first SN observation with signal-to-noise $\geqslant$ 5, we find that the gLSNe population with trailing images increases by $\sim$ 48\% with an average increase in brightness of $\sim$ 0.3 magnitudes (for a detailed breakdown by SN type, see Table \ref{tab:unknown_to_known}).

\section{Early Phase Supernovae Models} \label{sec: earlyPhaseSNe}

\begin{figure*}
    \centering
    \includegraphics[width=\textwidth]{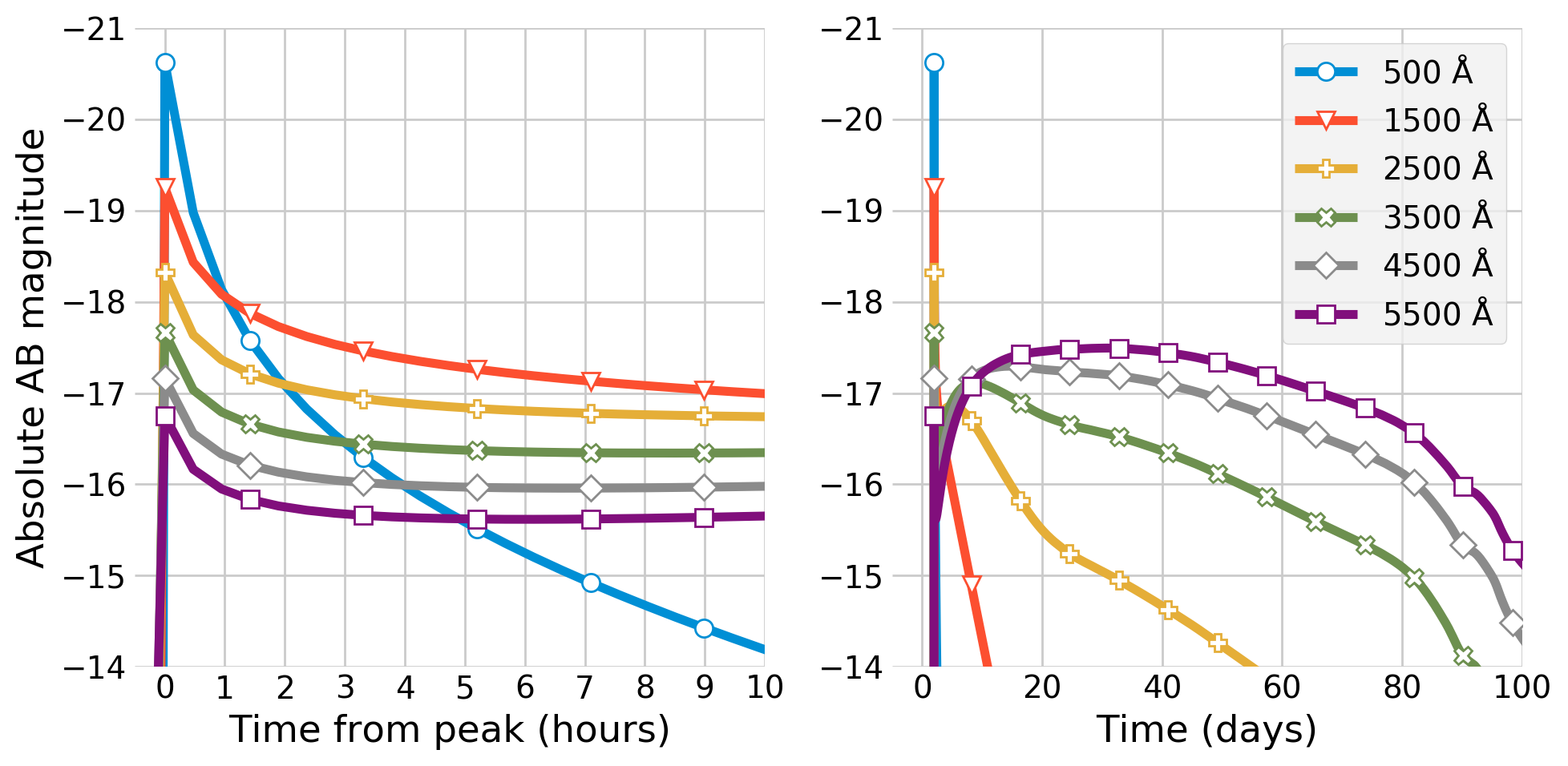}
    \caption{{Absolute AB magnitude for a Type IIP SN explosion as a function of rest frame wavelength. The light curve includes the initial shock breakout and was simulated using \texttt{SNEC} \citep[see][]{Morozova2015ApJ}. \textit{Left:} The initial hours of the light curve from the peak of the shock breakout, evolving over a timescale of $\sim$ 30 mins in the rest frame. \textit{Right:} The evolution of the full IIP light curve over 100 rest frame days.}}
    \label{fig:sb_MABvTime}
\end{figure*}

\begin{figure*}
    \centering
    \includegraphics[width=\textwidth]{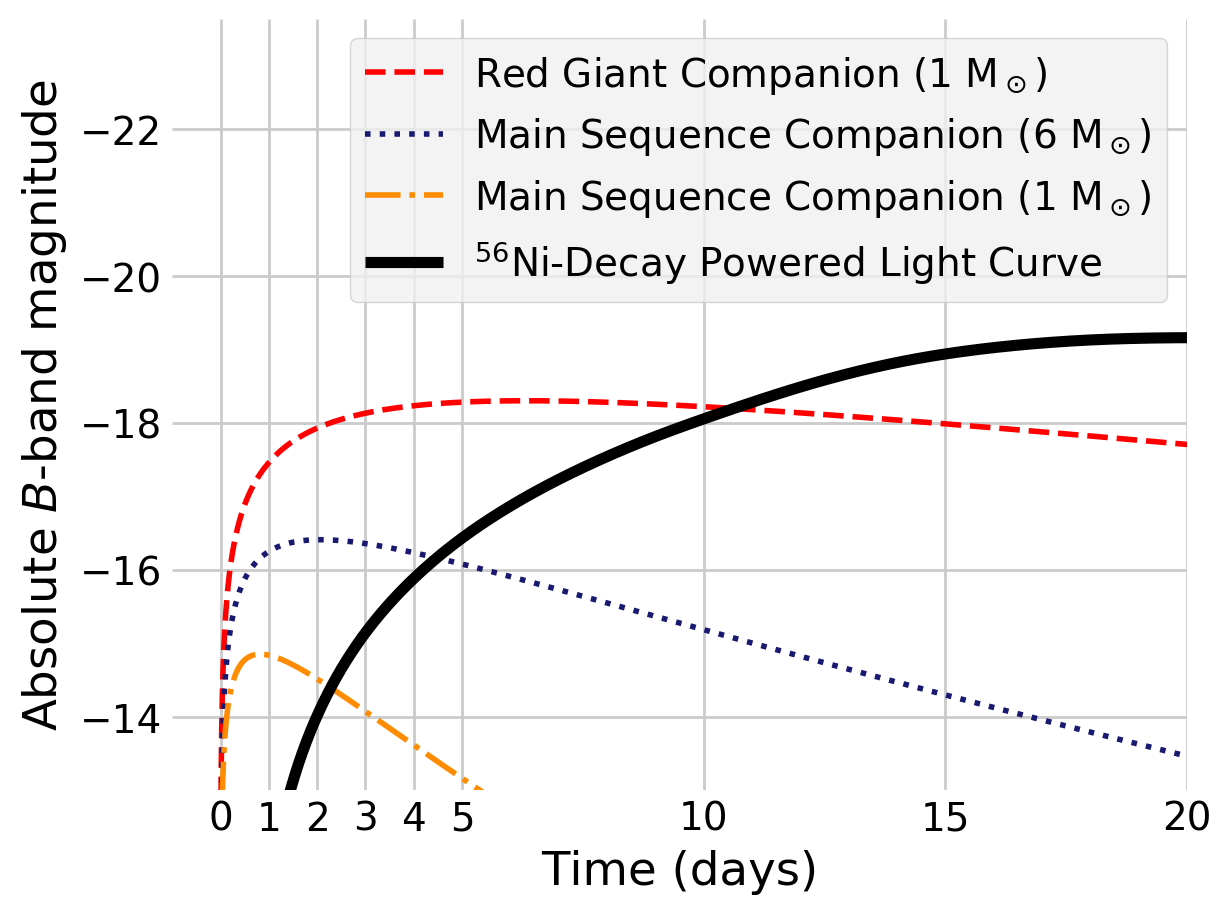}
    \caption{{Absolute B-band magnitude for a series of analytical companion shock cooling models from \citet{Kasen2010ApJ} plotted against a $^{56}$Ni-decay powered Type Ia SN light curve (derived from the \citealt{Hsiao2007ApJ_Ia} model, assuming a rise time of 20 rest-frame days from explosion and peak absolute $B$-band magnitude of -19.1). If there is a stellar companion, the observed flux during the earliest phases of the SN Ia will be dominated by the shock cooling component.}}
    \label{fig:IaCompanionLC}
\end{figure*}

\begin{figure}
    \centering
    \includegraphics[width=\columnwidth]{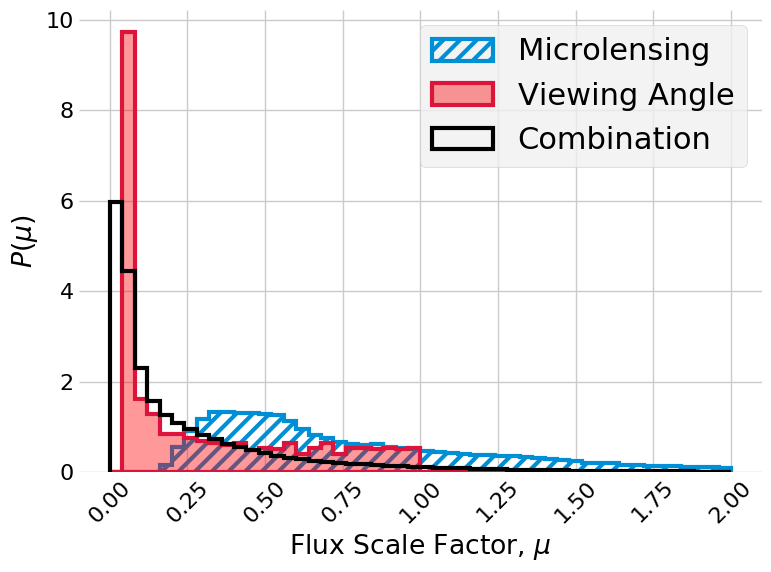}
    \caption{The effect of microlensing and viewing angle on the flux of a lensed SN image, relative to the case of no microlensing and directly viewing the shocked region. The microlensing effect averages to 1, but introduces scatter. The viewing angle introduces scatter and decreases the average flux by a factor of 0.3. The two effects are independent: black shows the convolution of the two effects.}
    \label{fig:microlensing_viewingangle_pdf}
\end{figure}

\begin{figure*}
    \centering
    \includegraphics[width=\textwidth]{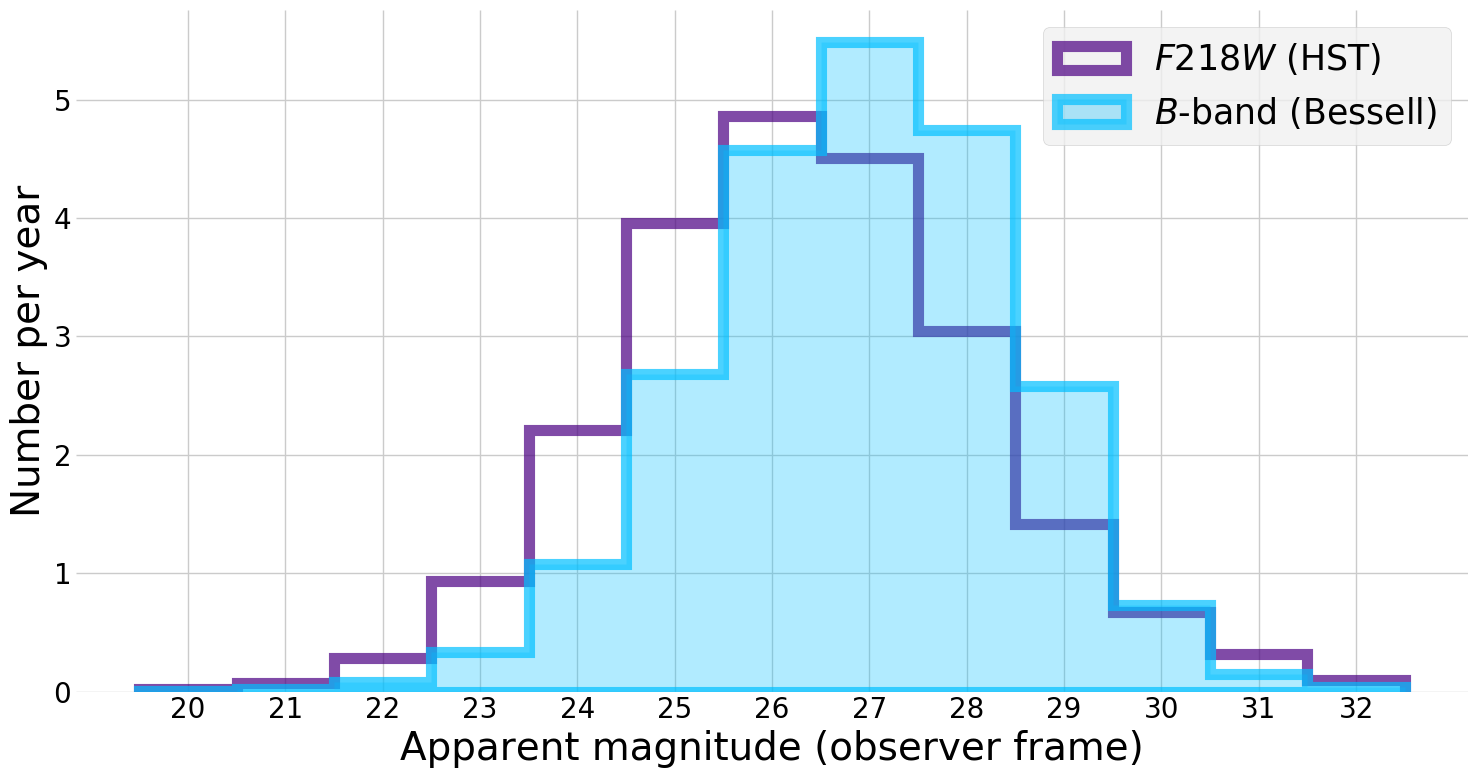}
    \caption{{Distribution of peak $B$-band and UV ($F218W$) observer frame magnitudes for a Type IIP shock breakout applied to the catalogue of trailing gLSNe IIP images.}}
    \label{fig:sbIIP_peakapparentmagnitude}
\end{figure*}

\begin{figure*}
    \centering
    \includegraphics[width=\textwidth]{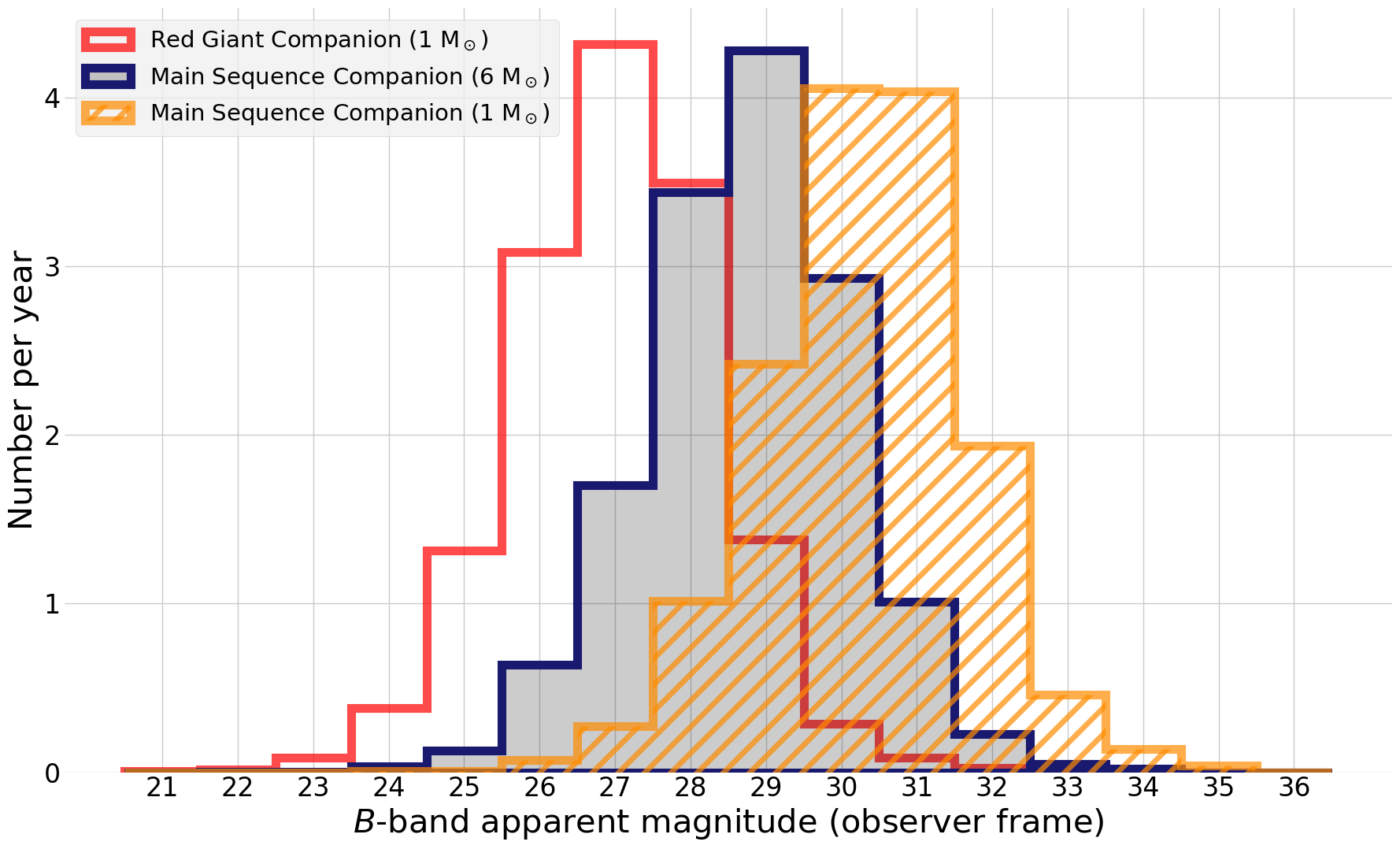}
    \caption{{Distribution of peak $B$-band observer frame magnitudes for Type Ia companion shock cooling events in the trailing images of Type Ia gLSNe, within one rest-frame day from explosion. We have performed the analysis across a series of plausible companion models. This figure assumes all Type Ia SNe in LSST are from the SD channel.}}
    \label{fig:sbIa_peakapparentmagnitude}
\end{figure*}

In this section of the paper, we apply light curves from a Type IIP detonation model (see Section \ref{subsec: IIPShockBreakout}) and a Type Ia SD companion cooling model (see Section \ref{subsec: IaCompanionCooling}) to the ensemble of LSST-discovered gLSN detailed in Section \ref{sec: Populations} in order to determine the early-phase peak brightness and rates of the SNe found in trailing gLSN images (including the effects of magnification and host galaxy/milky way extinction) and determine whether gLSNe can feasibly be used to observe early-phase SNe.

\subsection{Type IIP Shock Breakout} \label{subsec: IIPShockBreakout}
We model an instance of a Type IIP explosion using the SuperNova Explosion Code\footnote{\texttt{http://stellarcollapse.org/SNEC}} (\texttt{SNEC}), an open-source Lagrangian code for simulating the hydrodynamics and equilibrium-diffusion radiation transport in the expanding envelopes of SNe \citep[][]{Morozova2015ApJ}. For the progenitor star, we use the unstripped zero-age main sequence (ZAMS) reference star ($M_\mathrm{ZAMS} = 15 \ \mathrm{M_\odot}$) that was evolved by the open-source stellar evolution code \texttt{MESA} \citep{Paxton2011, Paxton2013} into a red supergiant with outer radius $R = 7.2 \times 10^{13}$ cm and total mass $M = 12.3 \ \mathrm{M_\odot}$\footnote{Some mass is lost in stellar winds during the star's evolution.}. We model the explosion as a black body and assume a constant grey opacity.

Figure \ref{fig:sb_MABvTime} shows the absolute magnitude of the Type IIP explosion over time, including the initial shock breakout, across a selection of wavelengths. The peak of the Type IIP shock breakout is brightest when observed at $\sim 400 \ \text{\AA}$ (extreme ultra-violet) in the source rest frame, with an absolute AB magnitude of $\sim$ -20.5. The rise and decline of the Type IIP shock breakout at early times is extremely rapid, occurring over a timescale of $\sim$ 30 minutes and is clearly distinct from the late-time light curve. The high-energy nature of the shock breakout means that the peak of the emission will be in the extreme UV in the source rest frame.

For strongly lensed images we must also account for microlensing by stars in the lensing galaxy in addition to the macromagnification from the entire lens galaxy. For sources much larger than the Einstein radius of a star, the granularity of the lens does not effect the total magnification of the source. This is not the case for gLSNe \citep{Foxley-Marrable2018}. Due to conservation of energy, microlensing by stars does not change the average magnification over an ensemble, but it can introduce significant scatter (\mbox{\citealt{Foxley-Marrable2018}}; \citealt{Goldstein2018ApJ}; \citealt{DoblerKeeton2006}). We use the microlensing magnification distributions from \citet{Vernardos2014, Vernardos2015} to build the probability density function for microlensing magnification. For simplicity sake, we assume all trailing images go through the region star field where 80 percent of the mass is in stars and 20\% in a smooth (dark matter) component. We assume all of the images have a lensing convergence and shear of 1.65, comparable to the typical values for trailing images found in Section \ref{sec: Populations}. The magnification distribution for such a microlensing configuration is shown in Figure \ref{fig:microlensing_viewingangle_pdf}. We assume the microlensing is achromatic at early times as found by \citet{Goldstein2018ApJ}, \citet{Huber2019} and \citet{Suyu2020}.

Figure \ref{fig:sbIIP_peakapparentmagnitude} shows the distribution of peak apparent magnitudes from applying our IIP shock breakout model to the LSST-discovered trailing gLSNe images, incorporating the effects of magnification (including microlensing by foreground stars) and extinction by dust (using the dust model of \citealt{GoldsteinNugentGoobar2019} and the reddening law of \citealt{Cardelli1989}). 

Assuming our model is representative of the IIP population, we predict to observe Type IIP shock breakouts at a rate of one per year at $\lesssim$ 24.1 mag in the $B$-band and $\lesssim$ 23.3 in the UV ($F218W$). However, since the shock breakout only lasts for $\sim$ 30 minutes, reaching this depth will require a large collecting aperture if spectroscopy or multiple points on the light curve are desired. 
Given that reappearance times will only typically be accurate to $2.6_{-1.4}^{+3.0}$ days for Type IIP gLSNe, a network of telescopes would be required to catch the shock breakout.

This result arises from the application of a single IIP detonation to the ensemble of Type IIP SNe from the \cite{GoldsteinNugentGoobar2019} catalogue. The absolute magnitudes of core collapse SNe can vary significantly, with a typical scatter of $\sim$ $\pm$1 mag for Type IIP SNe \citep{Li2011MNRAS, Richardson2014}. This variation in the magnitude of Type IIP SNe implies that our single realisation of the shock breakout is naive, and an ensemble of breakouts may shift, and will broaden the distribution of peak magnitudes shown in Figure \ref{fig:sbIIP_peakapparentmagnitude}.

\subsection{Type Ia Companion Shock Cooling} \label{subsec: IaCompanionCooling}
Using the analytic models from \cite{Kasen2010ApJ} we generate a series of shock cooling light curves for a non-degenerate companion star after shocking by the ejecta from a Type Ia SN. Radiative diffusion after shock-heating produces optical/UV emission. During the earliest phases of a SD Type Ia SN, the shock-heated emission is expected to exceed the radioactively powered luminosity {(see Figure \ref{fig:IaCompanionLC}, also refer to Figure 3 of \citealt{Kasen2010ApJ})}. Assuming a constant opacity and that the companion fills its Roche lobe, the luminosity and time scale for the shock cooling depends on the mass and stellar evolution stage of the companion. We investigate a 1 M$_\odot$ red giant (RG) companion, a 1 M$_\odot$ main sequence (MS) subgiant companion and a 6 M$_\odot$ MS subgiant companion.

The effect of viewing angle is such that companion shock cooling will on average be seen to be fainter than observing directly down onto the shocked region. However, back-scattering means that a few percent of the flux is observed even when observing from the opposite viewing angle to the shocked region \citep{Kasen2010ApJ}. To account for this effect, we assume the shocked region is described by a spherical cap on the surface of an opaque sphere. We assume that the cap has an opening angle of 15 degrees. The cap therefore covers $\sim$ 6 percent of the sphere. The relative flux observed as a function of viewing angle is proportional to the area of the cap projected onto a plane perpendicular to the viewing angle. We use the result of \citet{Urena2018} to perform the projection. The maximum flux is set to the analytic result of \citet{Kasen2010ApJ}. For viewing angles where the shocked region is occulted, we assume a minimum flux of 5 percent of the peak flux to account for back scattering. The flux scalings for the viewing angle effect are shown in Figure \ref{fig:microlensing_viewingangle_pdf}.

Figure \ref{fig:sbIa_peakapparentmagnitude} shows the range of peak observer $B$-band magnitudes for Type Ia companion shock cooling curves, predicted to be found in the trailing images of LSST gLSNe Ia, within one rest-frame day of explosion. If SNe Ia only came from the SD channel, we would expect to see at least one instance per year of shock cooling with a $B$-band magnitude of $\lesssim$ 26.3 assuming only 1 M$_\odot$ RG companions, $\lesssim$ 28.0 assuming only 6 M$_\odot$ MS companions and $\lesssim$ 29.6 assuming only 1 M$_\odot$ MS companions.

Since the shock cooling light curves evolve over a timescale of days (as opposed to minutes with the IIP shock breakout), the shock cooling can plausibly be caught with daily cadenced observations spread over the typical $3.3_{-1.4}^{+3.1}$ day time delay uncertainty for Type Ia gLSNe. 

On average, the brightness of sources in the $B$-band and the UV ($F218W$) are comparable to the $B$-band magnitude due to extinction by dust. However, if we are able to observe these sources in the UV, this would allow us to better differentiate between the very blue shock cooling light curve and the redder $^{56}$Ni driven light curve of the exploding WD \citep{Kasen2010ApJ}.

{In this section, we have only considered the possibility of early-time flux excess from ejecta-companion interaction i.e. from the SD channel. For example in the case of SN 2018oh, \citet{Dimitriadis2019} favoured the SD channel as the source of the early-time flux excess. However, \citet{Shappee2019ApJ} favoured the DD channel, noting that an off-center $^{56}$Ni distribution could produce a redder early-time flux component compared to the SD channel. Further analyses into sources of early-time flux other than ejecta-companion interaction will be left for future studies.}

\section{Constraining SNe Ia progenitor populations with early photometry} \label{sec: progenitorpop}
Observing companion shock cooling from a single SN Ia would be a demonstration that the SD channel is a viable progenitor system for producing SNe Ia. {However, it is plausible that the SN Ia population contains both SD and DD progenitors.} Observing - or not observing - shock cooling in a sample of SNe Ia can inform us about the progenitor population. 

If both the SD and DD channels are viable, the progenitor population should vary with redshift \citep{Childress2014}. The SD channel relies on Roche lobe overflow which happens at the end of the stellar main sequence life of the companion. The DD channel takes longer: both stars must evolve fully into WDs and then in-spiral due to loss of angular momentum through gravitational wave radiation. Thus the SD Ia population should approximately trace the cosmic star formation history, whereas there should be a longer delay between cosmic star formation and the explosions of DD SNe Ia \citep{Sullivan2006, Strolger2020}.

If the progenitor population varies as a function of redshift, it is of critical significance for Type Ia SN cosmology - if the mean magnitude of a SN Ia varies with redshift this will bias cosmological constraints derived assuming SNe Ia are standard candles.

In this section we investigate the ability of early time data to constrain the relative fraction of SD to DD populations, assuming the SD models follow the \citet{Kasen2010ApJ} shock cooling model and that DD Ia do not show early blue flux. The population of gLSNe Ia in Section \ref{sec: Populations}, the microlensing model in Section \ref{subsec: IIPShockBreakout} and the viewing angle model in Section \ref{subsec: IaCompanionCooling} give us a a probability density function for the amount of blue flux expected for each SD gLSN Ia. We test a toy model of progenitors where the ratio of SD to DD progenitors is $A$, and where all SD progenitors are 1 M$_\odot$ MS stars. 

The mathematics of this problem are akin to a coin flip experiment, except observational uncertainties mean that each `flip' is not uniquely identifiable as a SD or a DD and the SD model does not predict a unique value. The key probability theory is described in Appendix \ref{appendix:probability}.

\subsection{Constraining SN Ia progenitor populations with unlensed monitoring of the LSST deep drilling fields}
We first consider how well a blind survey could constrain the ratio of SD to DD progenitors, given a realistic observing strategy. LSST will observe 4 deep drilling fields every night for ten years with a total area of 38.4 square degrees. These fields will be observed $\sim$nightly in multiple filters, enabling high cadence photometry of early SN light curves without prior knowledge that a SN is about to occur. 

If the LSST deep drilling fields take $u$-band exposures every night to the ideal 5 sigma detection threshold of 23.9 \citep{Rothchild2019}, then LSST-deep would give nightly cadenced photometry of sufficient depth to observe shock cooling for 15 SNe per year, and 150 SNe over the 10 year duration of LSST, up to a limiting redshift limit of 0.115, assuming all SNe Ia are 1 M$_\odot$ MS subgaint companions, with optimal viewing angles (see Table \ref{tab:unlensedIaRates} for expected rates with limiting redshifts across all previously analysed companion models). The mean redshift of this population is 0.09.

\begin{figure*}
    \centering
    \includegraphics[width=\textwidth]{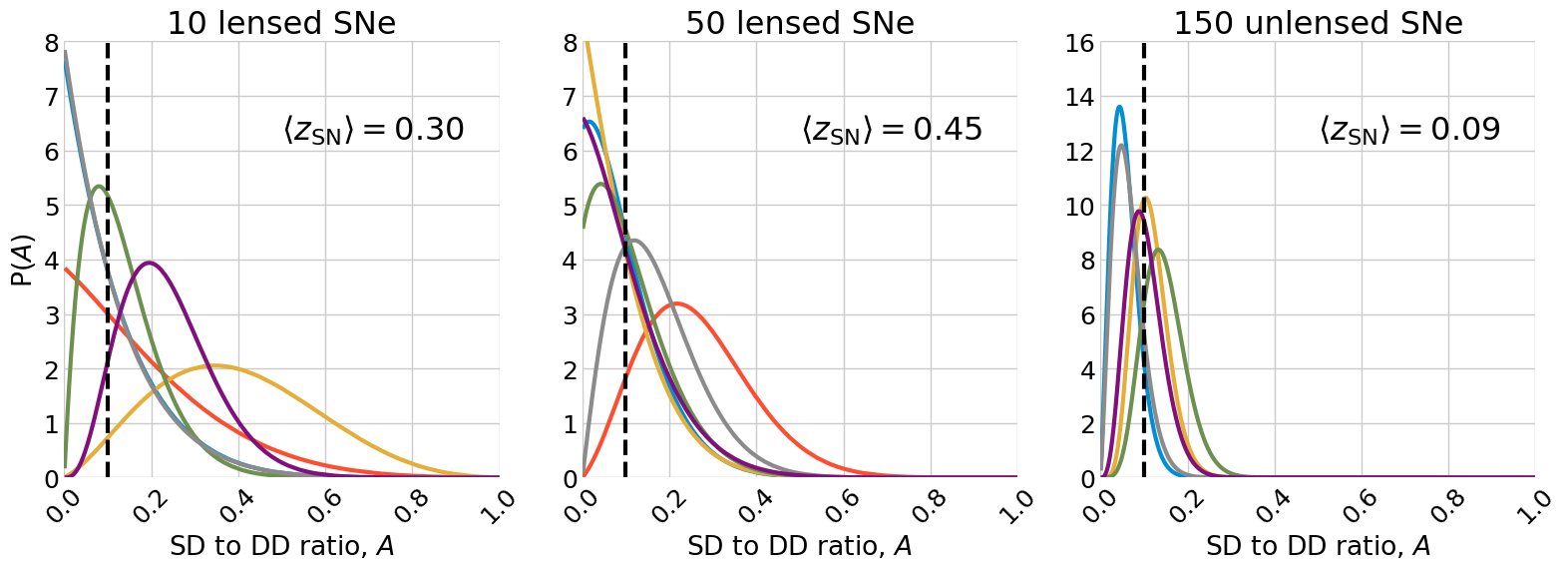}
    \caption{Forecast constraints on the ratio of SD to DD progenitors. From left to right: observations of the 10 best lensed trailing images with a 5 sigma depth of $m_B=28.7$, 50 lensed images to the same depth, and 150 unlensed images to $m_u=23.9$ assuming a blind search. Lines show the probability density function from a single realisation of the SN population, accounting for Poisson noise in the population, and randomness in the viewing angle, the SN redshifts and (for the lensed SNe only) magnification due to microlensing. Each PDF shows an equally likely realisation of the inferred P(SD/DD) given the assumed observing conditions. The input truth is shown by the dashed line.}
    \label{fig:PofA}
\end{figure*}

The forecast constraints on the ratio of SD to DD progenitors are stochastic, with the mean inferred value of $A$ and the error depending on shot noise in the realisations of the progenitor population, the realisations of the SN redshifts and the realisation of the viewing angles for the SD progenitors. We simulate 1000 realisations of 150 LSST SNe, assuming that ten percent of progenitors are SD ($A=0.1$; see \citealt{Livio2018}). Following the probability theory in Appendix \ref{appendix:probability}, we then infer $P(A)$ given the data in each realisation. We find that the 68 percent uncertainty on $A$ is $0.037 \pm 0.06$. The $P(A)$ inferred for 10 random realisations of this population is shown in Figure \ref{fig:PofA}.
When we assume there are no SD progenitors we find that the 95 percent upper limit on $A$ is $0.047 \pm 0.007$.

\subsection{Constraining SNe Ia progenitor populations with deep observations of LSST trailing images}
We now consider how well observations of the strongly lensed trailing images can be used to constrain the SN Ia progenitor population. As shown in Figures \ref{fig:superMegaLSST}, \ref{fig:superMegaLSSTIa} and \ref{fig:sbIa_peakapparentmagnitude}, the trailing images are at higher redshift and much fainter than can be observed during a single LSST exposure. However, the predictive power of lensing means that deep targeted follow-up is plausible. We assume a $B$-band 5 sigma depth of 28.7th magnitude, corresponding to a 60 minute exposure time on the European Extremely Large Telescope, with 0.8 arcsecond seeing and 7 days from new moon \citep{Liske2019}. 

If only a subset of the lenses can be followed up, focusing efforts on the brightest images minimises the uncertainty in $P(A)$. Because both the viewing angle and the microlensing effect are a priori unknown, it is impossible to predict which trailing images will show the brightest shock cooling events. However, an observer targeting the systems with the brightest trailing images as predicted by the macromodel will achieve the best signal to noise.

Assuming that ten percent of progenitors are SD ($A=0.1$), and that the 10 lens (macromodel predicted) brightest trailing images are followed up, we find that the  68 percent uncertainty on $A$ is $0.11_{-0.03}^{+0.04}$. The $P(A)$ inferred for 10 random realisations of this population is shown in Figure \ref{fig:PofA}. If the brightest 50 are followed up the uncertainty improves to $0.09 \pm 0.02$. Despite targeting 5 times more systems, there is only a modest improvement in uncertainty because most of these 50 are too faint for shock breakout to be detected even with a 5 sigma depth of 28.7 in the $B$-band unless there is significant microlensing magnification. 

When we assume there are no SD progenitors we find that the 95 percent upper limit on $A$ is $0.27 \pm 0.10$ and $0.19 \pm 0.05$ for followup of 10 and 50 lensed SNe respectively. 

Whilst the uncertainties for this lensed sample will be much larger than what a blind LSST deep drilling fields survey can achieve, the lensed sample is at higher redshift. The brightest 10 trailing images will come from SNe with a mean redshift of 0.3; for the brightest 50 it is 0.45.

\begin{table}
    \centering
    \begin{tabular}{p{2.5cm}p{2.5cm}p{1.5cm}}
    \toprule
    Companion model & LSST deep drilling field rates within $\mathrm{m_u} \lesssim 23.9$ (Year$^{-1}$) & Limiting redshift \\
    \midrule
    1 M$_\odot$ MS & 15   & 0.115 \\
    6 M$_\odot$ MS & 97  & 0.225 \\
    1 M$_\odot$ RG & 521 & 0.440 \\

    \bottomrule
    \end{tabular}
    \caption{Predicted rates for unlensed Type Ia shock cooling events to be observed in the LSST deep drilling fields, assuming the SN rates from \citet{Sullivan2006} and a limiting $u$-band magnitude of 23.9 from LSST.}
    \label{tab:unlensedIaRates}
\end{table}

\section{Time delay cosmology with early observations of lensed supernovae} \label{sec: cosmography}
Strong lensing time delays enable inference on cosmological parameters \citep{Refsdal1964}. However, measuring these time delays is observationally expensive \citep{Tewes2013}, requiring high cadence multi-season monitoring campaigns to yield robust time delays with several day precision. If observed in multiple images, the sharp features of an early phase gLSN would immediately provide a precise time-delay estimate. To do this would require identification of a quadruple imaged gLSNe before the explosion in at least 2 of the images. Across all SN types, LSST will discover $\sim$ 1 quad per year with multiple images remaining. Even if a sharp early phase feature were observed for every such system, this rate is too low to compete with the LSST sample of lensed AGN \citep{OguriMarshall2010, Liao2015}.

\section{Conclusions} \label{sec: Conc}

We have investigated the population of gLSNe systems which will be discovered in LSST and ZTF before the explosion occurs in the final image. We are now able to answer our initial questions:

\begin{enumerate}[]
    \item Will LSST and ZTF enable the discovery of gLSNe before the appearance of all multiple images?
\end{enumerate}

\noindent Across all SN types LSST is expected to find $\sim$ 110 trailing gLSNe per year, whilst ZTF will be finding significantly less at $\sim$ 1 trailing gLSNe per year (see Table \ref{tab:noPerYear} for a detailed breakdown). The LSST sample is dominated by doubles, whilst the ZTF sample is dominated by quads.

\begin{enumerate}[resume]
    \item How long is the time frame between the discovery of the system and explosion of the last image? How precisely can the last explosion time be predicted?
\end{enumerate}

\noindent Reaction times between discovery and the SN explosion in the final image are typically around $11.7^{+29.8}_{-9.3}$ days (Figures \ref{fig:superMegaLSST}e and \ref{fig:superMegaZTF}e). Assuming a 5\% precision on the time delay prediction from detailed lens modelling, {we find that we will be able to predict the reappearance of the SN in the final image to within $3.2_{-1.6}^{+3.4}$ days} (Figures \ref{fig:superMegaLSST}f and \ref{fig:superMegaZTF}f).

\begin{enumerate}[resume]
    \item How bright will the early phase light curves of Type IIP and Type Ia SNe found in the trailing images of LSST-discovered gLSNe get? 
\end{enumerate}
    
\noindent The vast majority of trailing images are demagnified by $\sim$ 1 or 2 magnitudes (Figure \ref{fig:deltamagImgno}), coupled with extinction by dust this will make obtaining early phase SN data using gLSNe an observationally challenging effort. 

For LSST gLSNe IIP, of order 1 trailing image per year will reach $\lesssim$ 24.1 in the $B$-band and $\lesssim$ 23.3 in the UV ($F218W$). Assuming the SD channel only for SNe Ia, we find that the LSST gLSNe population will include trailing images with one instance of a companion shock cooling emission per year in the $B$-band, with magnitude $\lesssim$ 26.3 assuming a 1 M$_\odot$ RG companion, $\lesssim$ 28.0 assuming a 6 M$_\odot$ MS subgiant companion and $\lesssim$ 29.6 assuming a 1 M$_\odot$ MS subgiant companion (Figure \ref{fig:sbIa_peakapparentmagnitude}).

\begin{enumerate}[resume]
    \item Can we use LSST-discovered gLSNe to make inferences on the progenitor population of Type Ia SNe with redshift? How will this compare with constraints from unlensed SNe Ia?
\end{enumerate}
    Figure \ref{fig:PofA} shows that assuming the brightest gLSN trailing images can be observed for 1 hour on the E-ELT the progenitor population can be constrained. When we assume there are no SD progenitors we find that the 95 percent upper limit on the fraction of 1 M$_\odot$ MS companions is $0.27 \pm 0.10$ and $0.19 \pm 0.05$ for followup of 10 and 50 lensed SNe respectively. Nightly $u$-band observations of the LSST deep drilling fields would yield more precise constraints, with 15 unlensed SNe per year bright enough to detect shock cooling from a 1 M$_\odot$ main sequence companion. Such observations would place a 5\% upper limit on the fraction of 1 M$_\odot$ main sequence companions at $\langle z \rangle = 0.09$. The gLSNe Ia are at higher redshifts, with even the 10 brightest systems having $\langle z \rangle = 0.30$.  Combining lensed and unlensed samples should constrain evolution in the Ia progenitor population and would place limits on progenitor evolution-induced systematics in Type-Ia SN cosmology.
    
\begin{enumerate}[resume]
    \item Can we measure precise time delays between the rapid early-phase light curves of gLSNe?
\end{enumerate}
    We find that this is unlikely to produce a cosmologically competitive sample of time delays. The rate of systems with multiple unexploded trailing images is below 1 per year even for LSST gLSNe.\\

In summary, during the LSST era catching the earliest phases of lensed SNe and constraining their progenitor physics is possible for Type Ia SNe if the community is willing to invest in deep ($\sim$ 26 to 30 mag in the $B$-band, depending on the progenitor) cadenced imaging for $3.2_{-1.6}^{+3.4}$ days either side of the predicted recurrence. 

\section*{Acknowledgements}
We thank Maria Vincenzi and the anonymous referee for constructive and meaningful discussions that were essential in the making of this paper. MF is supported by the University of Portsmouth, through a University Studentship. TC is supported by the Royal Astronomical Society through a Royal Astronomical Society Research Fellowship. Collaboration for this work was funded by a Royal Society International Exchange Grant (IE/170307).




\bibliographystyle{mnras}
\bibliography{ref} 


\appendix

\section{Probability theory for constraining two component progenitor populations}
\label{appendix:probability}

Constraining the underlying ratio of SD to DD SN Ia progenitors from an observed sample, is analogous to testing if a coin is fair given a finite number of flips. The mathematics of the progenitor problem is complicated slightly for two reasons: firstly, uncertainties in the observations mean that an individual observation does not perfectly discriminate between a SD and DD progenitor; secondly, whilst the DD is assumed to have no early blue flux the SD model does not predict a unique flux value due to viewing angle effects (and microlensing in the case of a strongly lensed SD Ia).

Assume a true population of Ia progenitors, where the underlying ratio of SD to DD progenitors is given by $A$. For any given supernova:
 \begin{equation}
P(SD|A) = A, P(DD|A)= 1 - A.
\end{equation}
Let us first consider the case where the data uniquely determines if the progenitor is a SD or DD. Let us denote $s$ as the number of SD and $d$ as the number of DD progenitors in a sample of $s+d$ events: 
\begin{equation}
P(s,d|A) \propto A^s (1-A)^{d}.
\end{equation}
Bayes theorem tells us that:
\begin{equation}
P(A|s,d)P(s,d)=P(s,d|A)P(A).
\end{equation}
Assuming a Uniform distribution for the prior on P(A) between 0 and 1 yields:
\begin{equation}
P(A|s,d) \propto A^s (1-A)^{d}.
\end{equation}

Let us now consider the case where the data does not uniquely determine if an event is SD or DD. 
For a single observation, $O_i$:
\begin{equation}
\begin{split}
P(O_i|A) &=P(O_i|SD)P(SD|A)+P(O_i|DD)P(DD|A)\\
&= A \times P(O_i|SD)+(1-A) \times P(O_i|DD).
\end{split}
\end{equation}
$P(DD|O_i|)=1-P(SD|O_i)$, are derived from the integral of the flux, $f$, predicted by the two models (a $\delta$ function at 0 for the DD model and a broader distribution for the SD model) over the window function consistent with the observed flux ($P(f|O_i)$):
\begin{equation}
P(SD|O_i)= \frac{\int_{-\infty}^{\infty} P(SD|f) P(f|O_i)  \mathrm{d}f}{\int_{-\infty}^{\infty} P(SD|f) \mathrm{d}f}.
\end{equation}
For multiple observations, $O$ the posterior is the product of the individual probabilities:
\begin{equation}
P(O|A) \propto \prod_{\forall i} \left(A \times P(O_i|SD)+(1-A) \times P(O_i|DD)\right),
\end{equation}
which can be inverted using Bayes theorem to infer P(A|O).

\section{LSST trailing gLSNe distributions by SN Type}
\begin{figure*}
    \centering
    \includegraphics[width=0.89\textwidth]{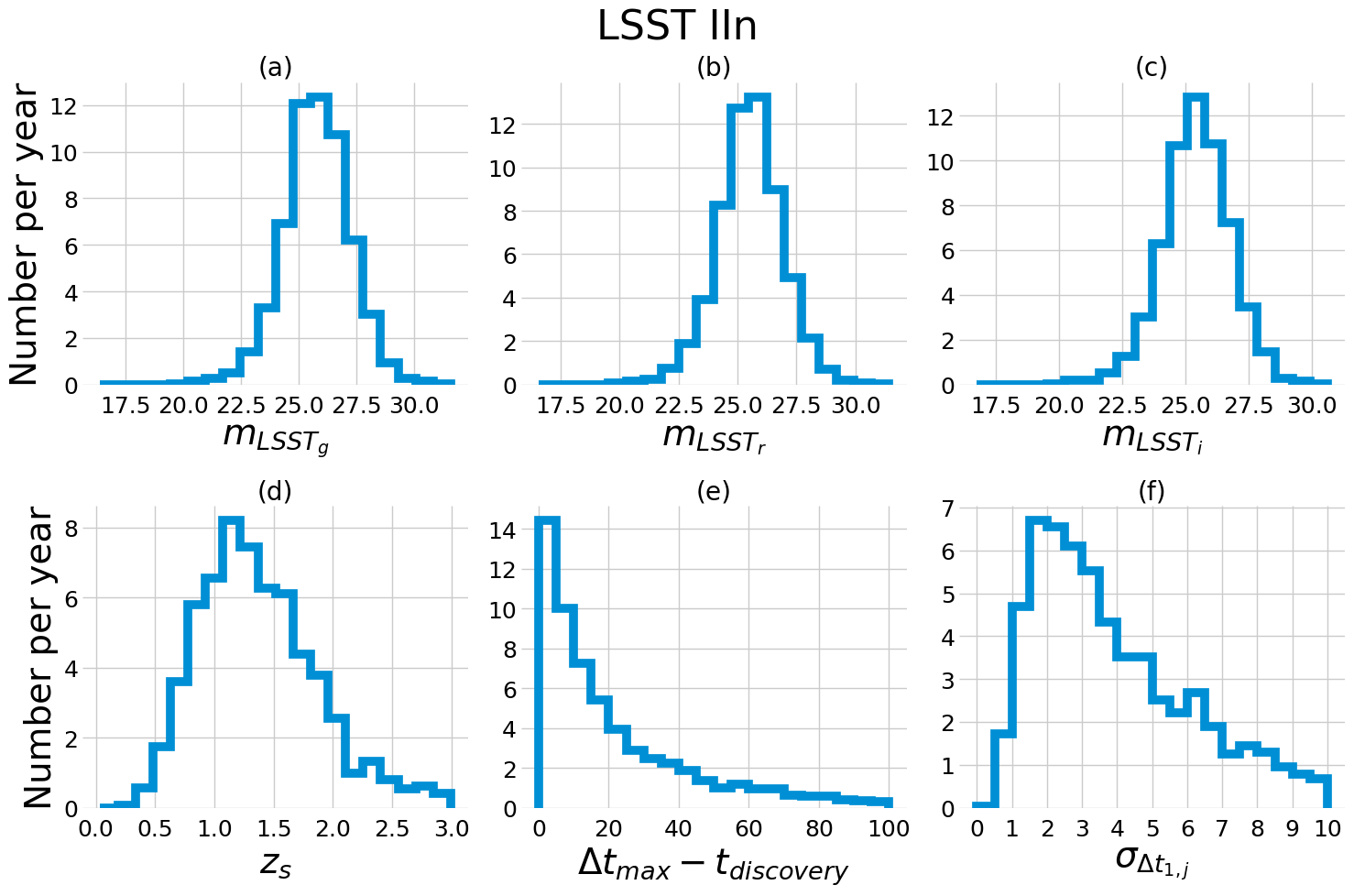}
    \caption{{Distributions and annual rates of LSST-discovered Type IIn gLSNe containing trailing images with unexploded SNe. See Table \ref{tab:superMegaFigDescriptions} for descriptions of the subplots.}}
    \label{fig:superMegaLSSTIIn}
\end{figure*}

\begin{figure*}
    \centering
    \includegraphics[width=0.89\textwidth]{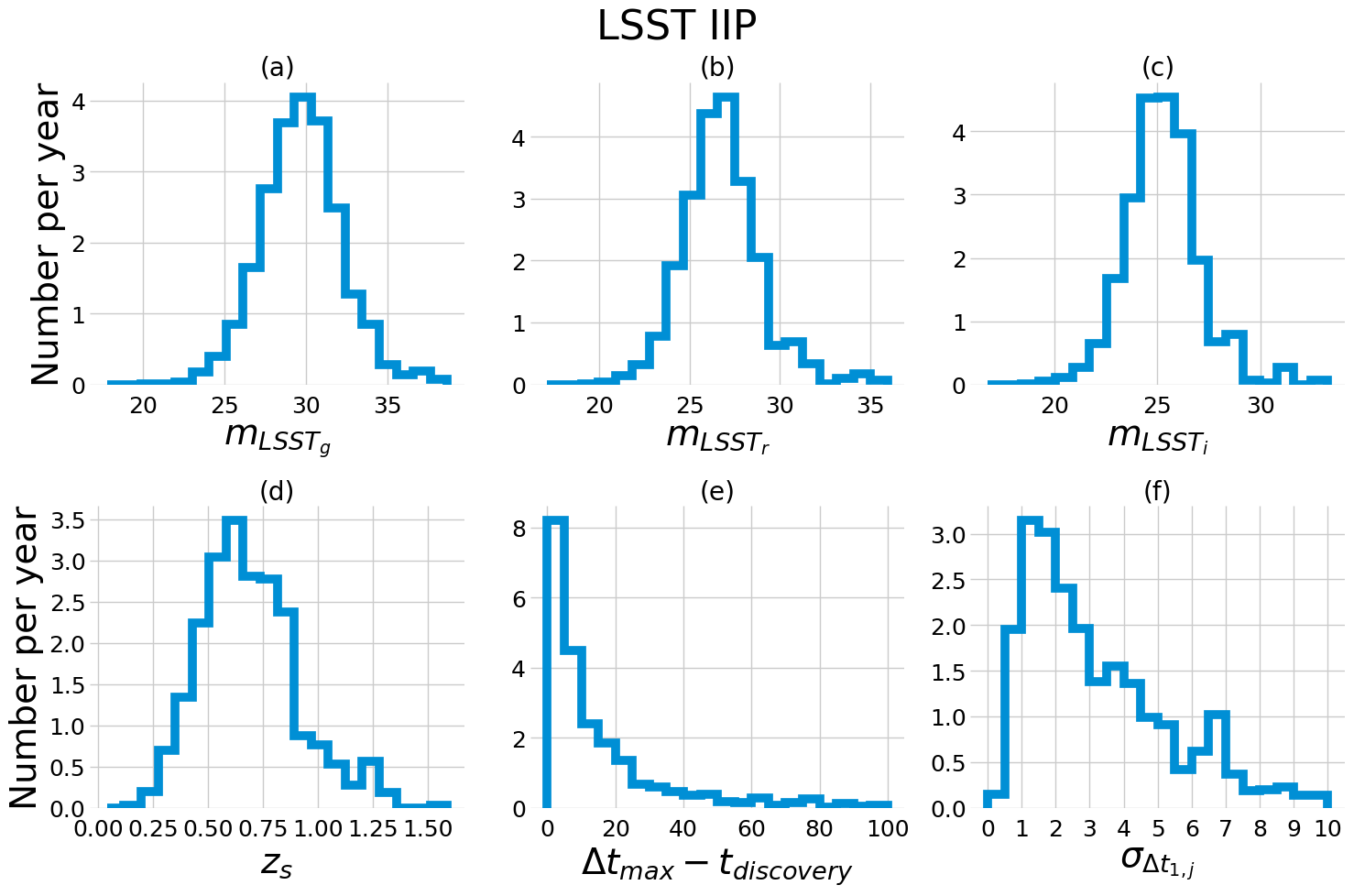}
    \caption{{Distributions and annual rates of LSST-discovered Type IIP gLSNe containing trailing images with unexploded SNe. See Table \ref{tab:superMegaFigDescriptions} for descriptions of the subplots.}}
    \label{fig:superMegaLSSTIIP}
\end{figure*}

\begin{figure*}
    \centering
    \includegraphics[width=0.89\textwidth]{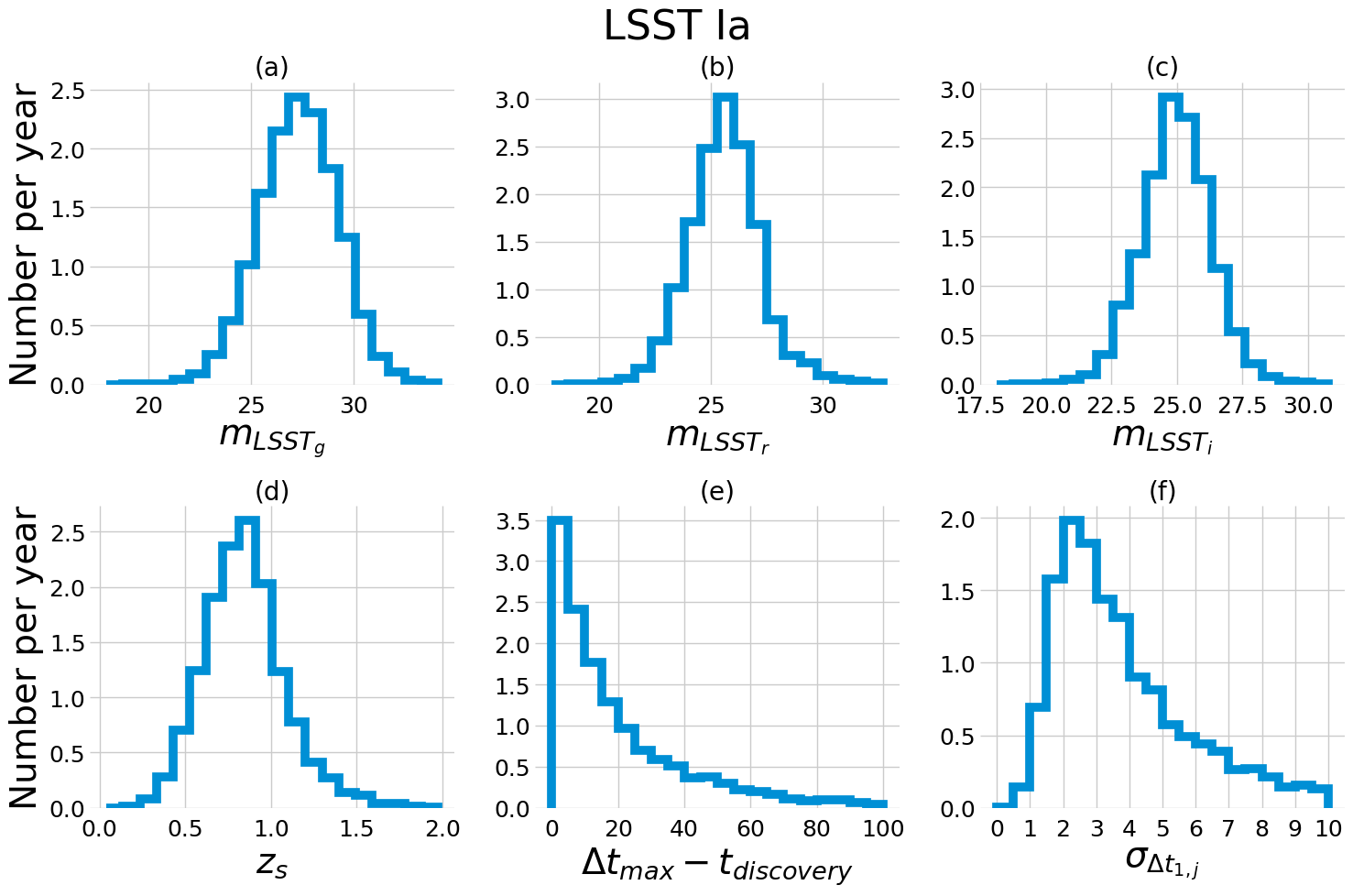}
    \caption{{Distributions and annual rates of LSST-discovered Type Ia gLSNe containing trailing images with unexploded SNe. See Table \ref{tab:superMegaFigDescriptions} for descriptions of the subplots.}}
    \label{fig:superMegaLSSTIa}
\end{figure*}

\begin{figure*}
    \centering
    \includegraphics[width=0.89\textwidth]{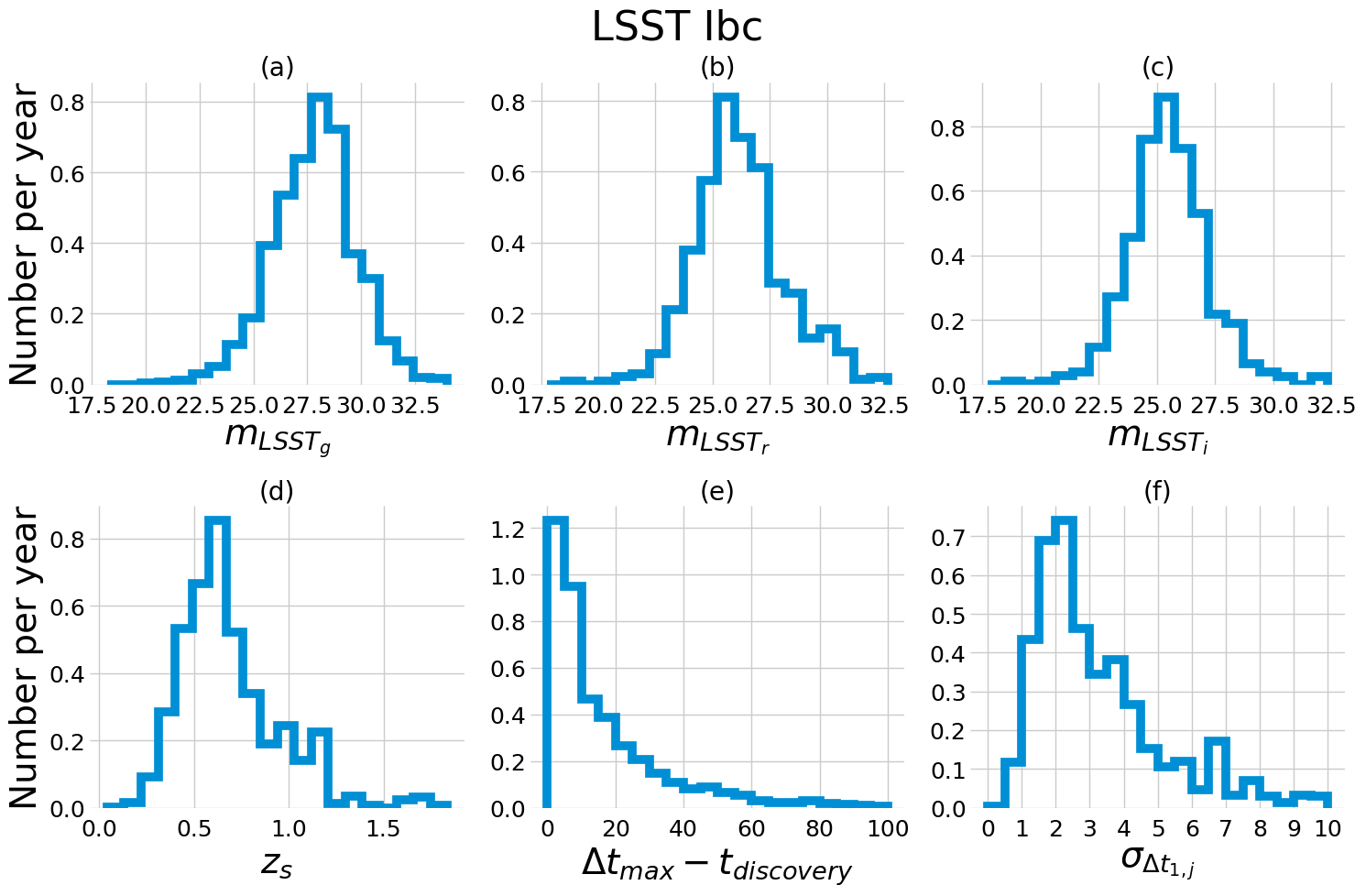}
    \caption{{Distributions and annual rates of LSST-discovered Type Ibc gLSNe containing trailing images with unexploded SNe. See Table \ref{tab:superMegaFigDescriptions} for descriptions of the subplots.}}
    \label{fig:superMegaLSSTIbc}
\end{figure*}

\begin{figure*}
    \centering
    \includegraphics[width=0.89\textwidth]{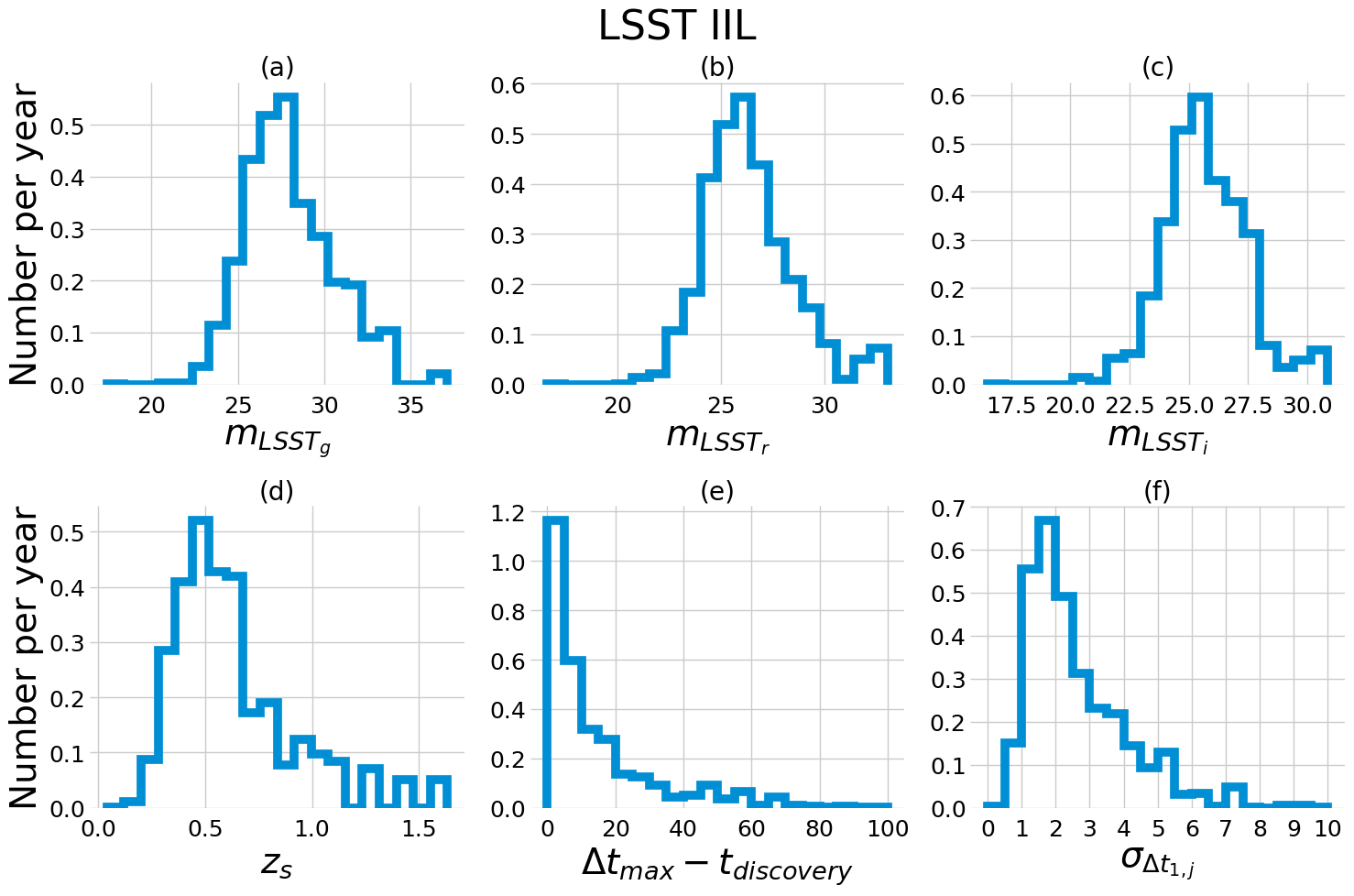}
    \caption{{Distributions and annual rates of LSST-discovered Type IIL gLSNe containing trailing images with unexploded SNe. See Table \ref{tab:superMegaFigDescriptions} for descriptions of the subplots.}}
    \label{fig:superMegaLSSTIIL}
\end{figure*}

\begin{figure*}
    \centering
    \includegraphics[width=0.89\textwidth]{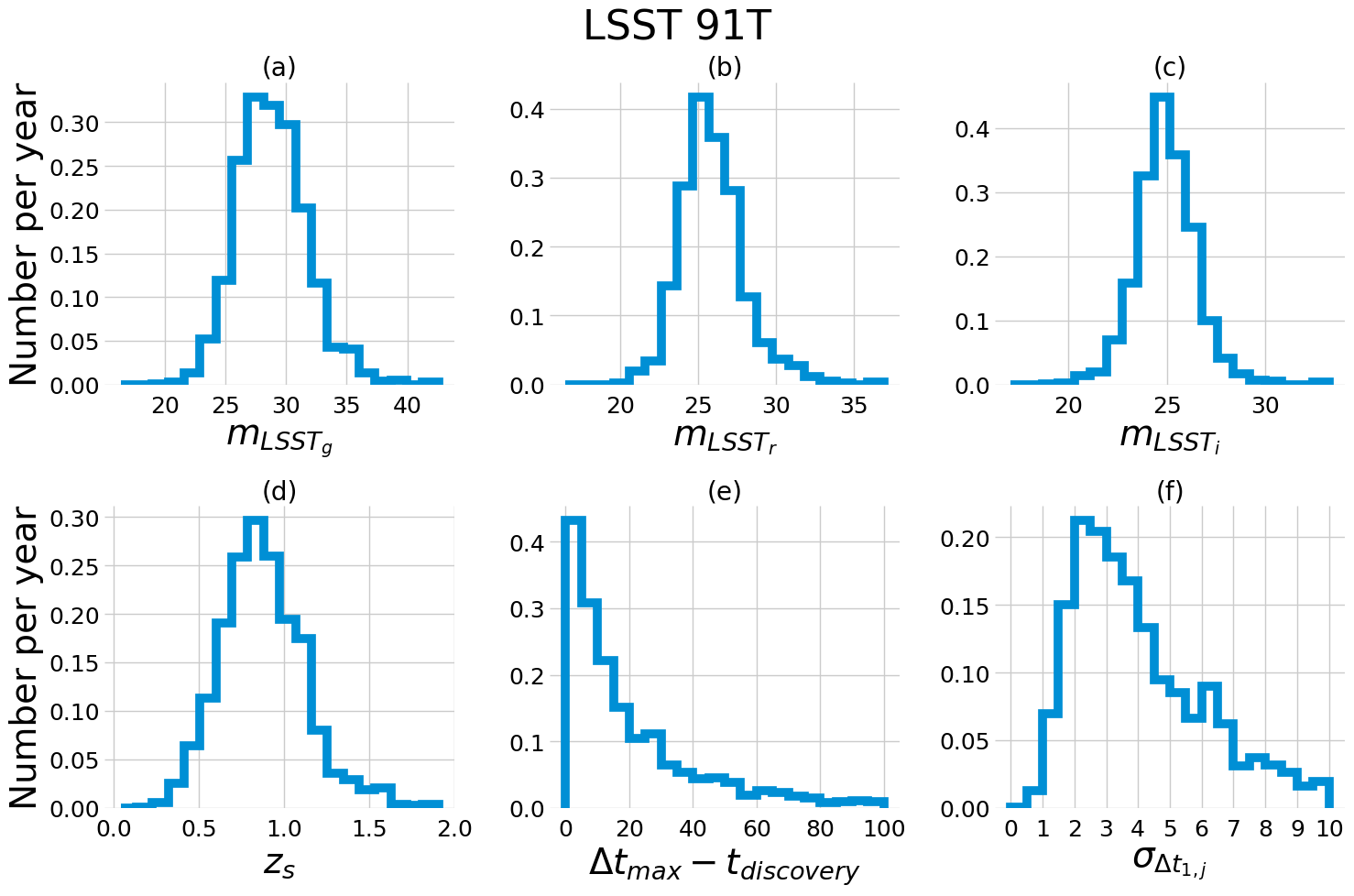}
    \caption{{Distributions and annual rates of LSST-discovered 91T-like gLSNe containing trailing images with unexploded SNe. See Table \ref{tab:superMegaFigDescriptions} for descriptions of the subplots.}}
    \label{fig:superMegaLSST91T}
\end{figure*}

\begin{figure*}
    \centering
    \includegraphics[width=0.89\textwidth]{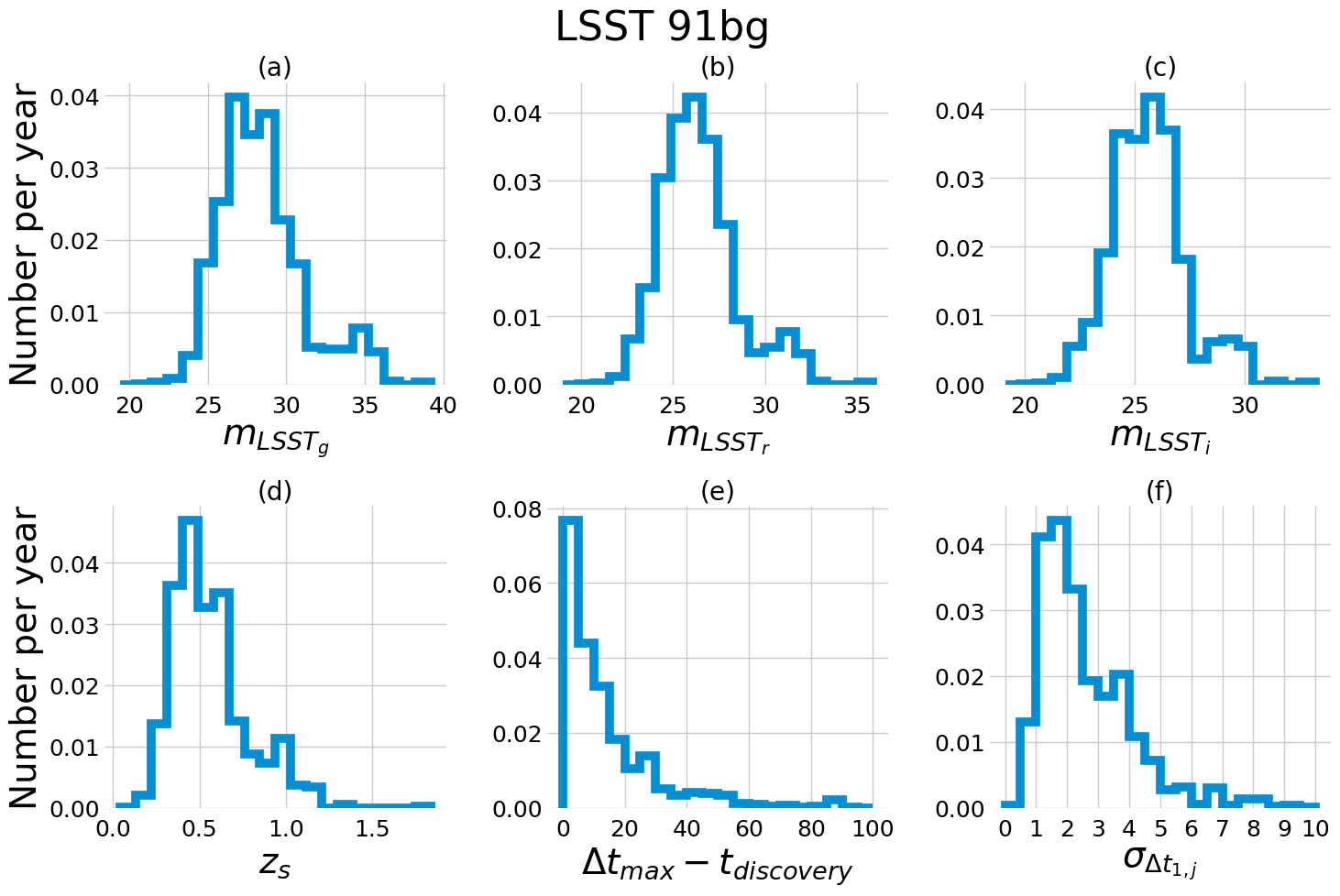}
    \caption{{Distributions and annual rates of LSST-discovered 91bg-like gLSNe containing trailing images with unexploded SNe. See Table \ref{tab:superMegaFigDescriptions} for descriptions of the subplots.}}
    \label{fig:superMegaLSST91bg}
\end{figure*}

We include the distributions and annual rates for LSST-discovered gLSNe categorised by SN type. We have not included ZTF due to low rates and sample size, resulting in some distributions being dominated by statistical noise. See figures \ref{fig:superMegaLSSTIIn} - \ref{fig:superMegaLSST91bg}.


\bsp	
\label{lastpage}
\end{document}